\documentclass[aps,pra,reprint,showpacs,amsmath,amssymb,10pt]{revtex4-1}
\usepackage{graphicx}% Include figure files
\usepackage{bm}% bold math
\usepackage{slashed}
\usepackage{color}

\begin{document}
\title{Frequency scaling law for nonlinear Compton and Thomson scattering:\\ Relevance of spin and polarization effects}
\author{K. Krajewska}
\email[E-mail address:\;]{Katarzyna.Krajewska@fuw.edu.pl}
\author{J. Z. Kami\'nski}
\affiliation{Institute of Theoretical Physics, Faculty of Physics, University of Warsaw, Pasteura 5,
02-093 Warszawa, Poland}
\date{\today}
\begin{abstract}
The distributions of Compton and Thomson radiation for a shaped laser pulse colliding with a free electron are calculated
in the framework of quantum and classical electrodynamics, respectively. We introduce a scaling law for the Compton and the Thomson 
frequency distributions which universally applies to long and short incident pulses. Thus, we extend the validity of
frequency scaling postulated in previous studies comparing nonlinear Compton and Thomson processes. The scaling law introduced
in this paper relates the Compton no-spin flipping process to the Thomson process over nearly the entire spectrum of emitted radiation,
including its high-energy portion. By applying the frequency scaling, we identify that both spin and polarization effects are responsible for differences between 
classical and quantum results. The same frequency scaling applies to angular distributions and to temporal power distributions of emitted radiation, which 
we illustrate numerically. 
\end{abstract}
\maketitle

\section{Introduction}

When an electron is scattered against a laser beam, an electromagnetic radiation is emitted; the process known as {\it Compton scattering} 
(for the most recent reviews, see, Refs.~\cite{review1,review2,review3,review4}). The complete theoretical description of this process is given in the framework of 
quantum electrodynamics (QED) by employing the Furry interaction picture~\cite{Furry} and by using the Volkov solutions~\cite{Volkov} in the initial and final electron states
(alternatively, if the process occurs in an underdense plasma, one can use the solutions derived in~\cite{Varro1,Varro2,Raicher1,Raicher2}).
In the low-energy limit, only classical aspects seem to play a role; the classical counterpart of the Compton scattering is known as {\it Thomson scattering}
(see, also Refs.~\cite{Lau,Umstadter}). In this case, the emitted radiation spectrum is obtained from the classical Newton-Lorentz equations, after substituting the resulting electron trajectory
in the Li\'{e}nard-Wiechert potentials~\cite{Jackson1975,LL2}. Both theoretical approaches shall be used in this paper assuming that the incident laser
beam can be modeled as a plane-wave-fronted pulse~\cite{Neville}.

The early works on nonlinear Compton~\cite{Brown,Goldman,Nikishov} and Thomson~\cite{Sengupta,Vaschaspati1,Vaschaspati2,Sarachik} scattering were based on a monochromatic plane wave approximation. 
A broad overview of the literature can be found in Refs.~\cite{review1,review2,review3,review4,Lau,Umstadter}. In the context of this paper, one should mention the paper by Heinzl 
{\it et al.}~\cite{Heinzl} who derived the scaling law relating the radiation spectra emitted in Compton and Thomson processes for the conditions relevant
to a definite number of photons; therefore describing a monochromatic incident field. While in Ref.~\cite{Heinzl} the frequency transformation concerned only 
backscattering in head-on geometry, in the following work~\cite{HSK} it was generalized for an arbitrary geometry allowing to
account, for instance, for finite size effects of detectors on the properties of emitted radiation. Further
comparison of Compton and Thomson spectra was performed in Refs.~\cite{Macken,Seipt,Mack,Boca2011} treating the case of 
a plane-wave-fronted pulse. The main features concerned the dependence of angular distributions 
of the emitted radiation on the carrier envelope phase of a driving pulse~\cite{Macken}, the blue shift of the classical 
energy spectrum, and the modification of the classical and quantum amplitudes~\cite{Seipt,Mack,Boca2011}. 
Specifically, in Ref.~\cite{Seipt} the scaling frequency law was introduced, for the conditions however that the notion of a number
of absorbed laser photons was still meaningful; therefore, describing a finite but sufficiently long driving pulse. In the present
paper, we shall further analyze the differences between classical and quantum results, with an emphasis on spin and polarization effects.
By introducing a frequency transformation, we identify the aforementioned effects to cause differences between quantum and classical results.
We show that, once these effects are accounted for, the scaling transformation introduced in this paper can be successfully applied to arbitrary laser pulses
(including short laser pulses, which is in contrast to the previous works). 
As we also demonstrate, our scaling law is applicable not only to frequency and angular distributions but 
also to temporal power distributions of emitted radiation. To our knowledge, the scaling of the latter has never been demonstrated before.

As we already mentioned, many of the existing calculations on nonlinear Compton and Thomson scattering treated the driving laser beam 
as a monochromatic plane wave (see, for instance, Refs.~\cite{Brown,Sengupta,Esarey,Ride,Faisal1,Faisal2,Faisal3,Gore,Panek,Ivanov,Harvey,Hartin,Popa1,Popa2}). Few works
on Compton scattering beyond this approximation can be found in literature~\cite{Narozhny,Boca2009,Macken,Seipt,Mack,Boca2011,Boca2012,KKcompton,KKpol}.
All of them concern a single electron response to the plane-wave-fronted pulse. Since a more accurate description of the scattering process
is accessible in the classical limit (see, for instance, Refs.~\cite{Browny,Gao,Krafft,Heinzl,Hart}), it is important to determine the relation
between quantum and classical calculations. This is particularly important in light of various applications of the Compton and
Thomson processes, including the production of ultra-short laser pulses in the x-ray domain~\cite{Esarey}, determining the
carrier envelope phase of intense ultra-short pulses~\cite{Macken}, measuring the electron beam parameters~\cite{Leemans},
and generating coherent comb structures in strong-field QED for radiation and matter waves~\cite{KKcomb}.

This paper is organized as follows. In Sec. II we introduce the theory of Compton scattering arising from quantum electrodynamics, 
whereas in Sec. III the same is done for Thomson scattering based on classical electrodynamics. In Sec. IV, the frequency scaling law
for emitted classical and quantum radiation is introduced. Sec. V contains numerical illustrations comparing classical and quantum energy spectra,
and discussing the validity of the introduced scaling law. This is done for long (Sec. VA) and short (Sec. VB) driving laser pulses,
and an emphasis is put on spin and polarization effects. In Sec. VI, we compare our results with the results of Ref.~\cite{Seipt},
postulating the frequency transformation between Compton and Thomson energy spectra induced by finite laser pulses. 
The analysis of the frequency scaling law is extended in Sec. VII to the angular distributions of generated radiation where we show how polarization vectors should be defined in order to achieve the agreement between the Compton and Thomson scattering. These investigations are supplemented in Sec. VIII by the discussion of the total energy of emitted radiation in the quantum and classical theories.
In Sec. IX, we illustrate that the same frequency scaling is applicable for the time-analysis of emitted radiation by these two processes. Our results are summarized in Sec. X.

\section{Compton scattering}

As in our previous investigations \cite{KKphase,KKcompton,KKbw,KKpol,KMK2013}, the laser pulse is assumed to be described by the vector potential
\begin{equation}
\bm{A}(\phi)=A_0 B[\bm{\varepsilon}_1 f_1(\phi)+\bm{\varepsilon}_2 f_2(\phi)],
\label{las1}
\end{equation}
where the shape functions $f_j(\phi)$ vanish for $\phi<0$ and $\phi>2\pi$. The duration of the laser pulse $T_{\mathrm{p}}$ introduces the fundamental frequency $\omega=2\pi/T_{\mathrm{p}}$ such that
\begin{equation}
\phi=k\cdot x=\omega \Bigl( t-\frac{\bm{n}\cdot\bm{r}}{c}\Bigr),
\label{las2}
\end{equation}
in which the unit vector $\bm{n}$ points in the direction of propagation of the laser pulse. We settle the real and orthogonal polarization 
vectors $\bm{\varepsilon}_j$, $j=1,2$ such that $\bm{n}=\bm{\varepsilon}_1\times\bm{\varepsilon}_2$. The constant $B>0$ is to be defined later. 
We also introduce the relativistically invariant parameter
\begin{equation}
\mu=\frac{|eA_0|}{m_{\mathrm{e}}c},
\label{las3}
\end{equation}
where $e=-|e|$ and $m_{\mathrm{e}}$ are the electron charge and mass.
With these notations, the electric and magnetic components of the laser pulse are equal to
\begin{equation}
\bm{\mathcal{E}}(\phi)=\frac{\omega m_{\mathrm{e}} c\mu}{e}B\bigl [\bm{\varepsilon}_1 f^{\prime}_1(\phi)+\bm{\varepsilon}_2 f^{\prime}_2(\phi)\bigr ],
\label{las4}
\end{equation}
and
\begin{equation}
\bm{\mathcal{B}}(\phi)=\frac{\omega m_{\mathrm{e}} c\mu}{ec}B\bigl [\bm{\varepsilon}_2 f^{\prime}_1(\phi)-\bm{\varepsilon}_1 f^{\prime}_2(\phi)\bigr ],
\label{las5}
\end{equation}
where '\textit{prime}' means the derivative with respect to $\phi$.

The shape functions are always normalized such that
\begin{equation}
\langle f_1^{\prime 2}\rangle +\langle f_2^{\prime 2}\rangle =\frac{1}{2},
\label{las6}
\end{equation}
where
\begin{equation}
\langle F \rangle=\frac{1}{2\pi}\int_0^{2\pi} F(\phi)\mathrm{d}\phi .
\label{las7}
\end{equation}
In our numerical illustrations, we shall choose the shape functions of the form
\begin{equation}
f(\phi)\propto\sin^2\Bigl(\frac{\phi}{2}\Bigr)\sin(N_{\mathrm{osc}}\phi).
\label{las8}
\end{equation}
Here, $N_{\mathrm{osc}}$ is the number of field oscillations within the pulse, therefore allowing one to define
the central laser frequency, $\omega_{\mathrm{L}}=N_{\mathrm{osc}}\omega$. In addition, we put the yet undetermined constant 
$B=N_{\mathrm{osc}}$, as we did in Ref.~\cite{KMK2013}.

When scattering a laser pulse off a free electron, a nonlaser photon is detected. It is described by the wave four-vector $K$ and, in the most general case, by the elliptically polarized four-vectors 
$\varepsilon_{{\bm K}\sigma}$ ($\sigma=1,2$) such that 
 \begin{equation}
K\cdot\varepsilon_{\bm{K}\sigma}=0,\quad \varepsilon_{\bm{K}\sigma}\cdot\varepsilon_{\bm{K}\sigma'}^*=-\delta_{\sigma\sigma'}.
\end{equation}
The wave four-vector $K$ satisfies the on-shell mass relation $K\cdot K=0$ as well as it defines the photon frequency $\omega_{\bm K}=cK^0=c|{\bm K}|$.
As shown in Ref.~\cite{KKbw}, $\varepsilon_{{\bm K}\sigma}$ can be chosen as the space-like vector, i.e., $\varepsilon_{{\bm K}\sigma}=(0,{\bm\varepsilon}_{{\bm K}\sigma})$.
The scattering is accompanied by the electron transition from the initial $({\rm i})$ to the final $({\rm f})$ state, each characterized by the four-momentum 
and the spin projection; $(p_{\rm i},\lambda_{\rm i})$ and $(p_{\rm f},\lambda_{\rm f})$. While moving in a laser pulse, the electron acquires an additional 
momentum shift~\cite{KKcompton} (see, also Ref.~\cite{KKbw}) which leads to a notion of the laser-dressed momentum:
\begin{align}
\bar{p}=p - & \mu m_\mathrm{e} c\Bigl(\frac{p\cdot\varepsilon_1}{p\cdot k}\langle f_1\rangle 
+ \frac{p\cdot\varepsilon_2}{p\cdot k}\langle f_2\rangle\Bigr)k \nonumber \\ 
+ & \frac{1}{2}(\mu m_\mathrm{e} c)^2\frac{\langle f_1^2\rangle+\langle f_2^2\rangle}{p\cdot k}k .  \label{com0}
\end{align}
It was discussed in Ref.~\cite{KKbw} that the dressed momenta defined according to Eq.~\eqref{com0}
are gauge-dependent, therefore they do not have clear physical meaning. Nevertheless, all formulas derived
in~\cite{KKcompton} depend on the quantity 
\begin{equation}
 P_N=\bar{p}_{\rm i}-\bar{p}_{\rm f}+Nk-K,
\label{com4}
\end{equation}
where the difference $\bar{p}_{\rm i}-\bar{p}_{\rm f}$ enters. This difference is already gauge-invariant and, as a consequence, all
quantities defined in~\cite{KKcompton} are as well. This concerns,
\begin{equation}
N_{\mathrm{eff}}=\frac{K^0+\bar{p}_{\mathrm{f}}^0-\bar{p}_{\mathrm{i}}^0}{k^0} =cT_{\mathrm{p}}\frac{K^0+\bar{p}_{\mathrm{f}}^0-\bar{p}_{\mathrm{i}}^0}{2\pi},
\label{cp12}
\end{equation}
which was proven to be also relativistically invariant~\cite{KKbw}.

We take the derivation of the Compton photon spectra from our previous paper~\cite{KKcompton}. As was presented there,
the frequency-angular distribution of energy of scattered photons for an unpolarized electron is given by the formula
\begin{equation}
\frac{\mathrm{d}^3E_{\mathrm{C}}}{\mathrm{d}\omega_{\bm{K}}\mathrm{d}^2\Omega_{\bm{K}}}=
\frac{1}{2}\sum_{\sigma=1,2}\sum_{\lambda_{\rm i}=\pm}\sum_{\lambda_{\rm f}=\pm}\frac{\mathrm{d}^3E_{\mathrm{C},\sigma}(\lambda_{\rm i},\lambda_{\rm f})}{\mathrm{d}\omega_{\bm{K}}\mathrm{d}^2\Omega_{\bm{K}}},
\label{com1}
\end{equation}
where
\begin{equation}
 \frac{\mathrm{d}^3E_{\mathrm{C},\sigma}(\lambda_{\rm i},\lambda_{\rm f})}{\mathrm{d}\omega_{\bm{K}}\mathrm{d}^2\Omega_{\bm{K}}}=\frac{e^2}{4\pi\varepsilon_0c}
\bigl|{\cal{A}}_{{\rm C},\sigma}(\lambda_{\rm i},\lambda_{\rm f})\bigr|^2
\label{com2}
\end{equation}
and the scattering amplitude equals
\begin{equation}
{\cal{A}}_{{\rm C},\sigma}(\lambda_{\rm i},\lambda_{\rm f})\!=\!\frac{m_{\rm e}cK^0}{2\pi\sqrt{p_{\rm i}^0k^0(k\cdot p_{\rm f})}}
\sum_N D_N\frac{1\!-\!{\rm e}^{-2\pi{\rm i}(N-N_{\mathrm{eff}})}}{{\rm i}(N-N_{\mathrm{eff}})}.
\label{com3}
\end{equation}
The scattering amplitude has been expressed as a Fourier series; for the coefficients $D_N$, the reader is referred to Eqs. (23) and (44) in Ref.~\cite{KKcompton}.
Note that, in contrast to a typical interpretation, integer indices $N$ in Eq.~(15) are not related to the number of emitted or absorbed laser photons or, in other words, 
to the period of a single field oscillation. They are related to the pulse duration, $T_{\mathrm{p}}$, or to the fundamental laser frequency, $\omega=2\pi/T_{\mathrm{p}}$.
For this reason, the respective Fourier expansion is meaningful for arbitrarily pulse durations. This is in contrast to Ref.~\cite{Seipt} where the expansion in
terms of a number of photons, thus characterized by the central laser frequency, $\omega_{\rm L}=N_{\rm osc}\omega$, was performed. The latter approach has clear
physical interpretation for relatively long driving pulses. At this point, we also recall that Eqs.~\eqref{com1},~\eqref{com2}, and~\eqref{com3} were derived 
using the conservation conditions: $P_N^-=0$ and ${\bm P}_N^\perp={\bm 0}$ (for more details, see Ref.~\cite{KKcompton}).
As we will explain in Sec. IV, these conditions are vital for deriving the Compton-Thomson frequency transformation.

\section{Thomson scattering}

In classical physics a point particle does not have a spin degree of freedom. Therefore, the description of nonlinear Thomson process 
introduced below applies to both bosons and fermions. At the moment, we assume that a particle possesses an arbitrary charge and mass, 
although at the end we shall apply this theory to electrons which have the smallest mass among charged particles.

Let a particle of charge $q$ and mass $m$ be accelerated from the initial time $t_{\mathrm{i}}$ to the final one $t_{\mathrm{f}}$. 
During this time interval it radiates, with the frequency-angular distribution of emitted energy given by the Thomson 
formula~\cite{Jackson1975} (we use the same notation for the radiation emitted during this process as for the Compton scattering)
\begin{equation}
\frac{\mathrm{d}^3E_{\mathrm{Th}}}{\mathrm{d}\omega_{\bm{K}}\mathrm{d}^2\Omega_{\bm{K}}}=
\frac{q^2}{4\pi\varepsilon_0 c}
\bigl | \bm{\mathcal{A}}_{\mathrm{Th}} \bigr |^2 ,
\label{thom1}
\end{equation}
where the vector amplitude is
\begin{equation}
\bm{\mathcal{A}}_{\mathrm{Th}}=\frac{1}{2\pi}\int_{t_{\mathrm{i}}}^{t_{\mathrm{f}}}
\bm{\Upsilon}(t)\exp\Bigl[\mathrm{i}\omega_{\bm{K}}\Bigl(t-\frac{\bm{n}_{\bm{K}}\cdot \bm{r}(t)}{c}\Bigr)\Bigr] \mathrm{d}t 
\label{thom2}
\end{equation}
and
\begin{equation}
\bm{\Upsilon}(t)=\frac{\bm{n}_{\bm{K}}\times [(\bm{n}_{\bm{K}}-\bm{\beta}(t))\times \dot{\bm{\beta}}(t)]}{\bigl(1-\bm{n}_{\bm{K}}\cdot \bm{\beta}(t)\bigr)^2}.
\label{thom3}
\end{equation}
Here the dot means the time derivative, $\bm{\beta}(t)=\dot{\bm{r}}(t)/c$ is the reduced velocity, and $\bm{n}_{\bm{K}}$ 
determines the direction of radiated energy with the polar and azimuthal angles, $\theta_{\bm{K}}$ and $\varphi_{\bm{K}}$, respectively.

In order to define the polarization properties of the Thomson radiation let us remark that for two polarization vectors 
$\bm{\varepsilon}_{\bm{K},\sigma}$ ($\sigma=1,2$) such that $\bm{\varepsilon}_{\bm{K},\sigma}\bot \bm{n}_{\bm{K}}$, one can write
\begin{equation}
\bm{\Upsilon}(t)=\bm{\varepsilon}_{\bm{K},1}(\bm{\varepsilon}_{\bm{K},1}^*\cdot \bm{\Upsilon}(t))+ \bm{\varepsilon}_{\bm{K},2}(\bm{\varepsilon}_{\bm{K},2}^*\cdot \bm{\Upsilon}(t)).
\label{thom4}
\end{equation}
Therefore,
\begin{equation}
\frac{\mathrm{d}^3E_{\mathrm{Th}}}{\mathrm{d}\omega_{\bm{K}}\mathrm{d}^2\Omega_{\bm{K}}}=
\sum_{\sigma=1,2}\frac{\mathrm{d}^3E_{\mathrm{Th},\sigma}}{\mathrm{d}\omega_{\bm{K}}\mathrm{d}^2\Omega_{\bm{K}}},
\label{thom5}
\end{equation}
where
\begin{equation}
\frac{\mathrm{d}^3E_{\mathrm{Th},\sigma}}{\mathrm{d}\omega_{\bm{K}}\mathrm{d}^2\Omega_{\bm{K}}}= \frac{q^2}{4\pi\varepsilon_0 c}
\bigl | \mathcal{A}_{\mathrm{Th},\sigma} \bigr |^2 
\label{thom6}
\end{equation}
and
\begin{equation}
\mathcal{A}_{\mathrm{Th},\sigma}=\bm{\varepsilon}_{\bm{K}\sigma}^*\cdot \bm{\mathcal{A}}_{\mathrm{Th}}.
\label{thom7}
\end{equation}
Eq. \eqref{thom6} determines the frequency-angular energy distribution of emitted radiation with polarization $\bm{\varepsilon}_{\bm{K}\sigma}$, 
which should be compared with the corresponding distribution, Eq.~\eqref{com2}, for the Compton scattering.

The acceleration $\bm{a}$ of a particle having charge $q$ and mass $m$ in arbitrary electric and magnetic fields, $\bm{\mathcal{E}}$ and $\bm{\mathcal{B}}$, 
is given by the formula \cite{LL2},
\begin{equation}
\bm{a}=\frac{q}{m}\sqrt{1-\bm{\beta}^2}\bigl[\bm{\mathcal{E}}-\bm{\beta}(\bm{\beta}\cdot\bm{\mathcal{E}}) +c\bm{\beta}\times\bm{\mathcal{B}}\bigr].  
\label{thom8}
\end{equation}
Therefore, the relativistic Newton-Lorentz equations, which determine the classical trajectory $\bm{r}(t)$ and the reduced velocity and acceleration, $\bm{\beta}(t)$ and $\dot{\bm{\beta}}(t)$, take the form
\begin{align}
\dot{\bm{r}}(t)=&c\bm{\beta}(t), \nonumber \\
\dot{\bm{\beta}}(t)=&\frac{qm_{\mathrm{e}}\omega\mu}{em}\sqrt{1-\bm{\beta}^2(t)} \nonumber \\
\times & \Bigl[ 
\bigl(\bm{\varepsilon}_1-\bm{\beta}(t)(\bm{\beta}(t)\cdot\bm{\varepsilon}_1)+\bm{\beta}(t)\times\bm{\varepsilon}_2\bigr)f^{\prime}_1(\phi) \nonumber \\
 &+ \bigl(\bm{\varepsilon}_2-\bm{\beta}(t)(\bm{\beta}(t)\cdot\bm{\varepsilon}_2)-\bm{\beta}(t)\times\bm{\varepsilon}_1\bigr)f^{\prime}_2(\phi)\Bigr] .
\label{thom9}
\end{align}
This is the system of ordinary differential equations that one has to solve with some initial conditions in order to calculate 
the Thomson distributions, Eqs. \eqref{thom1} or \eqref{thom6}. Without losing generality, we assume from now on that initially 
(at $t_{\mathrm{i}}=0$) the particle is at the origin of the coordinate system, $\bm{r}(0)=0$, with an arbitrary reduced velocity
such that $|\bm{\beta}(0)|<1$. Note that during the evolution, we have to determine not only the functions $\bm{r}(t)$, $\bm{\beta}(t)$, and $\dot{\bm{\beta}}(t)$, 
but also the finite time $t_{\mathrm{f}}$ after which the particle does not interact with the laser pulse, which means that the reduced acceleration vanishes. 
For presently available laser field intensities, this time can exceed the duration of the laser pulse, $T_{\mathrm{p}}$, by a few orders of magnitude 
which is due to the significant drift velocity in the pulse. Therefore, we have found it is more convenient to consider the phase $\phi$, 
instead of time $t$, as the independent variable of the Newton-Lorentz equations. In what follows, we solve the expanded system of equations 
\begin{align}
\frac{\mathrm{d}t(\phi)}{\mathrm{d}\phi}=&\frac{1}{\omega(1-\bm{n}\cdot\bm{\beta}(\phi))}, \label{thom9ex} \\
\frac{\mathrm{d}\bm{r}(\phi)}{\mathrm{d}\phi}=&\frac{c}{\omega}\frac{\bm{\beta}(\phi)}{1-\bm{n}\cdot\bm{\beta}(\phi)}, \nonumber \\
\frac{\mathrm{d}\bm{\beta}(\phi)}{\mathrm{d}\phi}=&\frac{qm_{\mathrm{e}}\mu}{em}\frac{\sqrt{1-\bm{\beta}^2(\phi)}}{1-\bm{n}\cdot\bm{\beta}(\phi)} \nonumber \\
\times & \Bigl[ 
\bigl(\bm{\varepsilon}_1-\bm{\beta}(\phi)(\bm{\beta}(\phi)\cdot\bm{\varepsilon}_1)+\bm{\beta}(\phi)\times\bm{\varepsilon}_2\bigr)f^{\prime}_1(\phi) \nonumber \\
 +& \bigl(\bm{\varepsilon}_2-\bm{\beta}(\phi)(\bm{\beta}(\phi)\cdot\bm{\varepsilon}_2)-\bm{\beta}(\phi)\times\bm{\varepsilon}_1\bigr)f^{\prime}_2(\phi)\Bigr] ,\nonumber
\end{align}
which also determines the dependence of time $t$ on the phase $\phi$. In this case,
\begin{align}
\bm{\mathcal{A}}_{\mathrm{Th}}=\frac{1}{2\pi}\int_{0}^{2\pi}
\bm{\Upsilon}(\phi) & \exp\Bigl[\mathrm{i}\frac{\omega_{\bm{K}}}{\omega}\phi \nonumber \\
+&\mathrm{i}\omega_{\bm{K}}\frac{(\bm{n}-\bm{n}_{\bm{K}})\cdot \bm{r}(\phi)}{c}\Bigr] \mathrm{d}\phi ,
\label{thom10}
\end{align}
and
\begin{equation}
\bm{\Upsilon}(\phi)=\frac{\bm{n}_{\bm{K}}\times [(\bm{n}_{\bm{K}}-\bm{\beta}(\phi))\times \bm{\beta}^{\prime}(\phi)]}{\bigl(1-\bm{n}_{\bm{K}}\cdot \bm{\beta}(\phi)\bigr)^2} ,
\label{thom11}
\end{equation}
where '\textit{prime}' means again the derivative with respect to the phase $\phi$. Similar modifications apply also to other formulas in this section.

In closing this section, let us note that in order to calculate the Thomson amplitude, Eq.~\eqref{thom10}, the first equation of the system~\eqref{thom9ex} 
is not necessary. It is included, however, to describe the classical trajectory and the reduced velocity and acceleration not only as functions of the phase $\phi$, 
but also as functions of the real time $t$. It appears that from the practical point of view such an expansion of the system of ordinary differential equations 
marginally increases the computational time. Moreover, it is well-known that the Newton-Lorentz equations with the electric and magnetic fields of the 
forms \eqref{las4} and \eqref{las5} can be solved in quadratures. However, this does not lead to significant simplifications as the numerical 
evaluation of integrals is equally time-consuming as the numerical solution of ordinary differential equations. Having this in mind, we choose to use
the current method.

\section{Frequency transformation}

Because in this section we discuss the quantum corrections to the frequency of emitted photons for the Compton process, exceptionally we restore here the Planck constant $\hbar$.

By inspecting Eq.~\eqref{com3} we find that the dominant contributions to the Compton amplitude come from such integer $N$'s that are very close to the real value $N_{\mathrm{eff}}$. 
This, along with the conservation conditions discussed following Eq.~\eqref{com3}, allow us to write down an approximate four momenta conservation condition
\begin{equation}
\bar{p}_{\mathrm{f}}=\bar{p}_{\mathrm{i}}+N_{\mathrm{eff}}\hbar k-\hbar K .
\label{fre1}
\end{equation}
Note that for very long laser pulses this equation is nearly exact for an integer $N_{\rm eff}$. However, for very short pulses it is fulfilled only approximately, 
which reflects the time-energy uncertainty relation. As for the Fermi's golden rule~\cite{BCK1992}, the above equation determines the most probable electron final momenta; 
only those momenta significantly contribute to the energy spectrum for which $N_{\mathrm{eff}}$ is as close as possible to an integer value.

The above equation determines the frequency of emitted Compton photon. Indeed, by squaring both sides of Eq.~\eqref{fre1} and after some algebra we arrive at
\begin{equation}
\omega_{\bm{K}}=\frac{N_{\mathrm{eff}}c k\cdot p_{\mathrm{i}}}{q_{\mathrm{i}}\cdot n_{\bm{K}}+N_{\mathrm{eff}}\hbar k\cdot n_{\bm{K}}} ,
\label{fre2}
\end{equation}
where the four-vector $q_{\mathrm{i}}$ equals
\begin{equation}
q_{\mathrm{i}}=\bar{p}_{\mathrm{i}}+\mu m_{\mathrm{e}}c (\langle f_1\rangle \varepsilon_1+\langle f_2\rangle \varepsilon_2 ) ,
\label{fre2a}
\end{equation}
and represents the gauge-invariant dressing of the initial momentum $p_{\mathrm{i}}$ (see, Ref.~\cite{KKbw}).

In the classical limit ($\hbar\rightarrow0$), we obtain from Eq.~\eqref{fre2} the frequency, which we denote by $\omega_{\bm{K}}^{\mathrm{Th}}$ and attribute to the classical Thomson frequency,
\begin{equation}
\omega_{\bm{K}}^{\mathrm{Th}}=\frac{N_{\mathrm{eff}}c k\cdot p_{\mathrm{i}}}{q_{\mathrm{i}}\cdot n_{\bm{K}}} .
\label{fre3}
\end{equation}
In both formulas, Eqs.~\eqref{fre2} and \eqref{fre3}, there is still an unknown real number $N_{\mathrm{eff}}$, which can be eliminated by expressing 
$\omega_{\bm{K}}$ by $\omega_{\bm{K}}^{\mathrm{Th}}$. In doing so, we define the \textit{cut-off} frequency
\begin{equation}
\omega_{\mathrm{cut}}=\frac{c}{\hbar}\frac{n\cdot p_{\mathrm{i}}}{n\cdot n_{\bm{K}}} .
\label{fre4}
\end{equation}
This quantity has a purely kinematic character. Namely, it depends only on the geometry of the process and, except for the direction of propagation of the pulse, 
it is independent of the laser field parameters responsible for the dynamical aspects of the process. With this definition we find that
\begin{equation}
\omega_{\bm{K}}=\frac{\omega_{\bm{K}}^{\mathrm{Th}}}{1+\omega_{\bm{K}}^{\mathrm{Th}}/\omega_{\mathrm{cut}}} ,
\label{fre5}
\end{equation}
or
\begin{equation}
\omega_{\bm{K}}^{\mathrm{Th}}=\frac{\omega_{\bm{K}}}{1-\omega_{\bm{K}}/\omega_{\mathrm{cut}}} .
\label{fre6}
\end{equation}
As it follows from the Thomson theory, the frequency of the generated radiation can be arbitrary large. On the other hand, for the quantum Compton process the frequency must fulfill the boundaries \cite{KKcompton}
\begin{equation}
0 < \omega_{\bm{K}} < \omega_{\mathrm{cut}},
\label{fre7}
\end{equation}
at least for an arbitrary laser pulse for which the plane-wave-fronted approximation applies. 
Eqs.~\eqref{fre5} and \eqref{fre6} exactly reflect these properties of classical and quantum radiation
which, together with the numerical analysis presented below, justify the interpretation of $\omega_{\bm{K}}^{\mathrm{Th}}$ 
as the frequency generated by the classical process. These relations can be put in the relativistically covariant form for the wave four-vectors,
\begin{equation}
K^{\mathrm{Th}}=\nu K, \quad \frac{1}{\nu}=1-\hbar\frac{k\cdot K}{k\cdot p_{\mathrm{i}}}.
\label{fre7a}
\end{equation}

The discussion presented above leads to the common interpretation of the validity of the Thomson theory. It states that the results coincide 
with the ones derived from the Compton theory provided that
\begin{equation}
\omega_{\bm{K}} \ll \omega_{\mathrm{cut}}.
\label{fre8}
\end{equation}
For instance, in the reference frame of the initial electrons it adopts the form
\begin{equation}
\omega_{\bm{K}} \ll \frac{m_{\mathrm{e}}c^2}{\hbar}\frac{1}{1-\cos\theta_{\bm{K}}},
\label{fre9}
\end{equation}
where $\theta_{\bm{K}}$ is the angle between the direction of the laser field propagation and the direction of emission of Compton photons. 
This shows that the Thomson theory could be valid even for Compton photons of energy comparable to or larger than $m_{\mathrm{e}}c^2$, provided 
that the emission angle $\theta_{\bm{K}}$ is sufficiently small. Note that the above validity condition is independent of the intensity of the 
laser field. Does it mean that we could apply the classical theory to arbitrarily intense laser pulses? The answer to this question is, in 
our opinion, unknown since both classical and quantum theories have been derived from the lowest order of perturbation theory.
For the Thomson theory we have neglected the radiation reaction effects, whereas for the Compton theory we have disregarded the radiative corrections 
to the leading Feynman diagram.

\section{Numerical analysis}
\begin{figure}
\includegraphics[width=7.0cm]{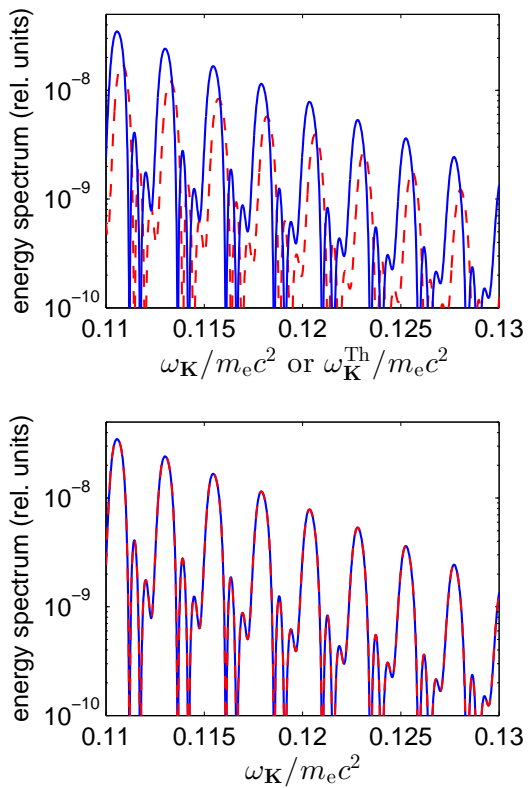}%
\caption{(Color online) Energy spectra for the Compton scattering (solid blue line), Eq.~\eqref{com1}, and for the Thomson scattering (dashed red line), 
Eq.~\eqref{thom5}, for the linearly polarized laser field propagating in the $z$-direction with the polarization vector along the $x$-axis. The laser 
pulse parameters are: $\mu=1$, $N_{\mathrm{osc}}=32$, and $\omega_{\mathrm{L}}=3\times 10^{-6}m_{\mathrm{e}}c^2$. The scattered 
radiation is linearly polarized in the scattering plane, and it is characterized by the polar and azimuthal angles, $\theta_{\bm{K}}=0.98\pi$ 
and $\varphi_{\bm{K}}=0$, respectively. The initial electron propagates in the opposite direction with respect to the $z$-axis, with momentum 
$|\bm{p}_{\mathrm{i}}|=50m_{\mathrm{e}}c$. In the upper panel, the Thomson spectrum is calculated for the frequency $\omega_{\bm{K}}^{\mathrm{Th}}$. 
In the lower panel, this frequency is transformed to the Compton frequency $\omega_{\bm{K}}$ by applying the scaling law \eqref{fre5} 
and, in addition, the Thomson energy spectrum is multiplied by 2. For these particular parameters, $\omega_{\mathrm{cut}}\approx 50m_{\mathrm{e}}c^2$.
\label{shiftct20130629a}}
\end{figure}
\begin{figure}
\includegraphics[width=7.0cm]{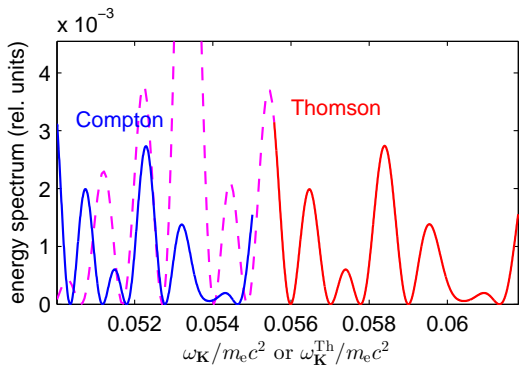}%
\caption{(Color online) Energy spectra for the Compton scattering (solid blue line for the no-spin-flipping process, $\lambda_{\mathrm{i}}\lambda_{\mathrm{f}}=1$), 
Eq.~\eqref{com2}, and for the Thomson scattering (dashed magenta and solid red lines), Eq.~\eqref{thom5}. The driving pulse propagates in the $z$-direction
and is linearly polarized along the $x$-axis. The remaining laser field parameters are such that $\mu=10$, $N_{\mathrm{osc}}=16$, and $\omega_{\mathrm{L}}=0.3 m_{\mathrm{e}}c^2$. 
The direction of scattered radiation is given by the polar and azimuthal angles, $\theta_{\bm{K}}=0.99\pi$ and $\varphi_{\bm{K}}=0$, respectively. 
These parameters are specified in the rest frame of incident electrons. The Thomson spectrum, multiplied by the factor 0.9, is calculated for the frequency 
$\omega_{\bm{K}}^{\mathrm{Th}}$. In this reference frame and for these parameters, $\omega_{\mathrm{cut}}\approx m_{\mathrm{e}}c^2/2$.
\label{ctcheck3ff}}
\end{figure}

In the following, the laser field propagation is chosen in the $z$-direction, and the electron spin degrees 
of freedom are defined with respect to this axis. We introduce a notion of the {\it scattering plane} which is determined by the propagation direction of the incident pulse
and the emitted radiation, thus defining the $(xz)$-plane. For an incident laser field, we choose the shape function $f_1(\phi)$ as a sine-squared function~\eqref{las8} whereas $f_2(\phi)=0$
[see, Eq.~\eqref{las1}]. Also, it is assumed that ${\bm\varepsilon}_1={\bm e}_x$ and ${\bm\varepsilon}_2={\bm e}_y$ in Eq.~\eqref{las1}.

We start our numerical analysis for the parameters, presented in the caption to Fig.~\ref{shiftct20130629a}, for which one can expect 
the agreement between both theories. The presented frequency range of emitted radiation is much smaller than the cut-off frequency, $\omega_{\mathrm{cut}}$. 
The quantum Compton distribution [Eq.~\eqref{com1}] is calculated as a function of frequency $\omega_{\bm{K}}$, whereas the classical Thomson distribution
[Eq.~\eqref{thom5}] as a function of $\omega_{\bm{K}}^{\mathrm{Th}}$. The comparison of the two is shown in the upper panel. We see that the spectra 
are very similar except that the classical one is blue-shifted with respect to its quantum equivalent, and that both differ in amplitude. This was realized in the previous papers~\cite{Boca2011,Seipt,Mack}.
However, if we present the classical distribution such that its frequency $\omega_{\bm{K}}^{\mathrm{Th}}$ is scaled to $\omega_{\bm{K}}$, according to Eq.~\eqref{fre5}, we get the agreement between 
these two distributions. The agreement is up to a multiplicative factor which, for the whole range of the considered frequencies, is roughly equal to 2. This result suggests the following scaling law:
\begin{equation}
\frac{\mathrm{d}^3E_{\mathrm{C},\sigma}}{\mathrm{d}\omega_{\bm{K}}\mathrm{d}^2\Omega_{\bm{K}}} = \gamma(\omega_{\bm{K}},\Omega_{\bm{K}})
\frac{\mathrm{d}^3E_{\mathrm{Th},\sigma}}{\mathrm{d}\omega_{\bm{K}}^{\mathrm{Th}}\mathrm{d}^2\Omega_{\bm{K}}} \bigg|_{\omega_{\bm{K}}^{\mathrm{Th}}=\frac{\omega_{\bm{K}}}{1-\omega_{\bm{K}}/\omega_{\mathrm{cut}}}} .
\label{num1}
\end{equation}
As we mentioned, the frequency transformation, Eq.~\eqref{fre5}, has a purely geometric origin. On the other hand, the differences between the quantum 
and classical dynamics for these processes are hidden in the pre-factor, $\gamma(\omega_{\bm{K}},\Omega_{\bm{K}})$, which is unknown; it appears,
however, from our numerical analysis that it is a smooth function of its arguments, as compared to the Compton and Thomson distributions that are, in general, rapidly
changing functions. For this reason, in a frequency interval containing a few oscillations of these distributions, one can write that
\begin{equation}
\frac{\mathrm{d}^3E_{\mathrm{C},\sigma}}{\mathrm{d}\omega_{\bm{K}}\mathrm{d}^2\Omega_{\bm{K}}} \sim
\frac{\mathrm{d}^3E_{\mathrm{Th},\sigma}}{\mathrm{d}\omega_{\bm{K}}^{\mathrm{Th}}\mathrm{d}^2\Omega_{\bm{K}}} \bigg|_{\omega_{\bm{K}}^{\mathrm{Th}}=\frac{\omega_{\bm{K}}}{1-\omega_{\bm{K}}/\omega_{\mathrm{cut}}}} .
\label{num1a}
\end{equation}
This means that the Compton and Thomson theories give similar results in the sense that after rescaling the Thomson frequency and multiplying 
the Thomson distribution by a constant factor both distributions become almost identical. This is illustrated in the lower panel of Fig.~\ref{shiftct20130629a}.

As mentioned above, the exact form of the factor $\gamma(\omega_{\bm{K}},\Omega_{\bm{K}})$ in Eq.~\eqref{num1} is not known.
Our numerical analysis shows, however, that for $\omega_{\bm{K}}\ll\omega_{\mathrm{cut}}$ it is nearly equal to 1, whereas for other values of
$\omega_{\bm{K}}$ (even those close to $\omega_{\mathrm{cut}}$, where the applicability of the classical approach is questionable) 
$\gamma(\omega_{\bm{K}},\Omega_{\bm{K}})$ is a slowly varying function of its arguments. These properties enable for a fast theoretical analysis of spectral and temporal 
characteristics of radiation generated during the interaction of electrons with intense laser pulses. Namely, in order 
to determine these properties, one has to apply a rather complicated and numerically demanding formalism of the strong-field QED. 
For sufficiently intense laser pulses, such an analysis becomes extremely time-consuming as distributions of generated radiation 
are very rapidly oscillating functions. This means that, in order to determine them properly, one has to perform 
the calculation of quantum probability amplitudes for densely distributed sample points. The scaling law allows to speed up this 
procedure significantly, with some limitations concerning polarization properties of emitted radiation and spin dynamics of 
electrons interacting with strong laser pulses, as it is going to be discussed below. Indeed, to determine the slowly changing 
factor  $\gamma(\omega_{\bm{K}},\Omega_{\bm{K}})$ it is sufficient to calculate quantum and classical amplitudes for sparsely 
distributed sample points. Having determined $\gamma(\omega_{\bm{K}},\Omega_{\bm{K}})$, one can perform computationally much less time-consuming (although not fully 
appropriate for very intense laser pulses, as it will follow shortly from our analysis) classical investigations of the process.
These results multiplied by $\gamma(\omega_{\bm{K}},\Omega_{\bm{K}})$ give a good estimation of quantum distributions, 
that can be compared afterward with experimental results. Let us also remark that another possibility of determining 
approximately $\gamma(\omega_{\bm{K}},\Omega_{\bm{K}})$ (however, in our opinion less accurate) has been suggested in Ref.~\cite{Seipt}. 
It consists in relating this factor to the ratio of the corresponding distributions for the monochromatic plane wave (see, Eq.~(59) in Ref.~\cite{Seipt}).

%\begin{widetext}
\begin{figure*}
\includegraphics[width=12.0cm]{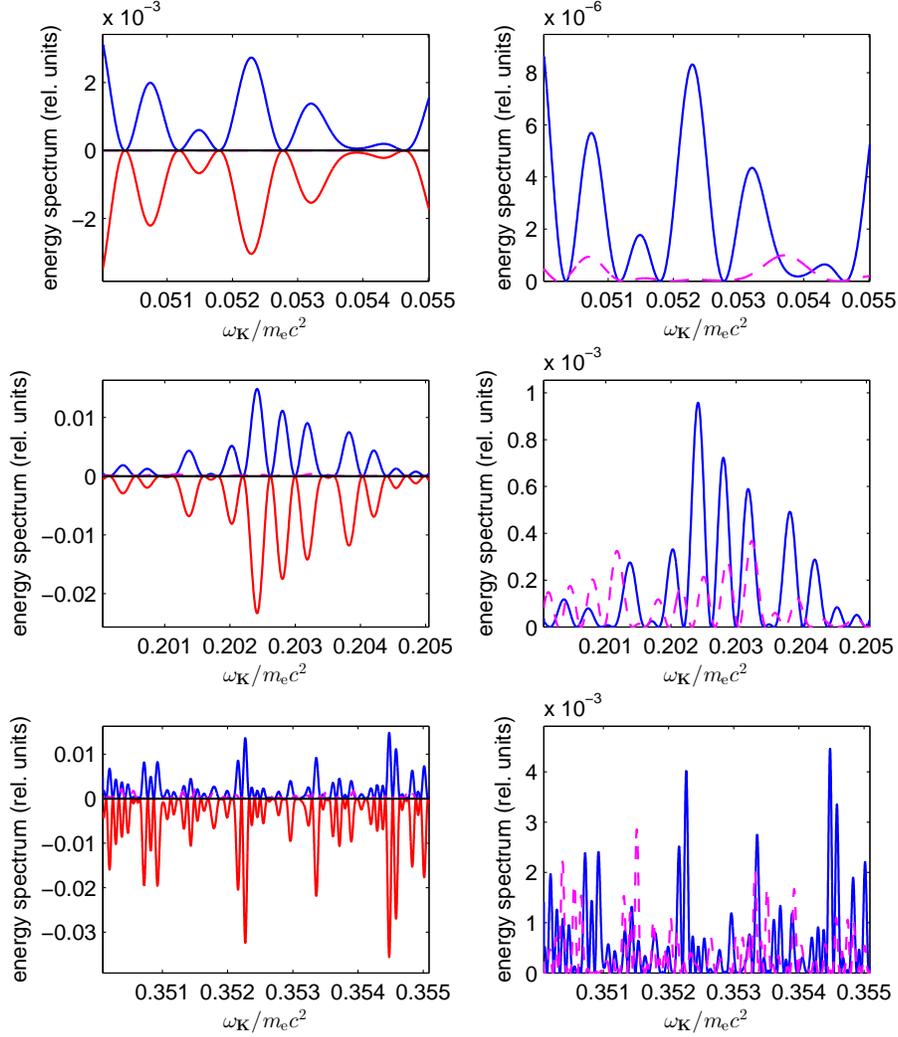}%
\caption{(Color online) Energy spectra for the Compton scattering (solid blue line for the no-spin-flipping process, $\lambda_{\mathrm{i}}\lambda_{\mathrm{f}}=1$, 
dashed magenta line for the spin-flipping process, $\lambda_{\mathrm{i}}\lambda_{\mathrm{f}}=-1$), Eq.~\eqref{com2}, and for the Thomson scattering (solid red line, 
reflected with respect to the horizontal black line), Eq.~\eqref{thom5}, and for the same parameters as in Fig.~\ref{ctcheck3ff}. In the left column, 
the energy spectra are presented for emitted radiation polarized linearly in the scattering plane for three chosen frequency domains. The right column 
displays the energy spectra for perpendicularly polarized emitted radiation, for which the Thomson theory gives 0. The Thomson spectrum is calculated 
for the frequency $\omega_{\bm{K}}^{\mathrm{Th}}$ and then the frequency is transformed to the Compton frequency $\omega_{\bm{K}}$ by applying the scaling law \eqref{fre5}. 
For these particular parameters and for the reference frame considered, $\omega_{\mathrm{cut}}\approx m_{\mathrm{e}}c^2/2$.
\label{ctcheck3a}}
\end{figure*}
%\end{widetext}

Note that the Compton scattering has a much richer structure than its classical counterpart. First of all, it depends on the electron 
spin degrees of freedom. Moreover, if the laser pulse is linearly polarized in the scattering plane the Thomson theory predicts no radiation 
with polarization perpendicular to this plane, which is in contrast to the Compton theory. (For more works on polarization effects in Thomson
scattering, we refer the reader to Refs.~\cite{Popa1,Popa2,Krafft,Boca2011}; the polarization effects in Compton scattering were analyzed more closely
in Refs.~\cite{KKpol,Ivanov,King}.) The agreement between both theories occurs when, for Compton scattering, 
the spin-flipping processes as well as the emission of radiation polarized perpendicularly to the scattering plane take place with small probabilities. 
For this reason, the frequency scaling law has to be more specific. In the following, we shall demonstrate that, as long as the classical theory
predicts the emission of radiation, its distribution is similar to the quantum one for spin-conserved processes.

Since the Compton and Thomson theories are relativistically invariant, in the remaining part of this paper we restrict our numerical analysis to the reference frame 
of the incident electron. 

\subsection{Long laser pulses}

For long laser pulses, the four-momentum conservation condition, Eq.~\eqref{fre1}, is well satisfied with significant probability amplitudes only for an integer $N_{\mathrm{eff}}$. 
Therefore, let us consider the long pulse with $N_{\mathrm{osc}}=16$. In Fig.~\ref{ctcheck3ff}, we present the respective Compton energy spectrum for 
$0.1\leqslant \omega_{\bm{K}}/\omega_{\mathrm{cut}}\leqslant 0.11$, and the Thomson one for $\omega_{\bm{K}}^{\mathrm{Th}}$ changing over a wider interval. 
The Thomson distribution is represented by the dashed magenta line but part of it, which is similar to the Compton one for the spin no-flipping channels, 
is covered by the continuous red line. These two parts of the distributions are similar in the sense that, by applying the scaling transformation~\eqref{fre6} 
and by multiplying the Thomson distribution by the factor $\gamma(\omega_{\bm{K}},\Omega_{\bm{K}})=0.9$, both solid lines (the red and the blue one) coincide. 
Note that the similar parts of the quantum and classical distributions are from the frequency domains which are separated from each other. Below, we  
show that, even though such a separation can be very large, both theories give similar results.

To this end we compare in Fig.~\ref{ctcheck3a} these two distributions in more detail. This is done for the same laser pulse parameters 
but for three different frequency domains: $0.1\leqslant \omega_{\bm{K}}/\omega_{\mathrm{cut}}\leqslant 0.11$ (top row), 
$0.4\leqslant \omega_{\bm{K}}/\omega_{\mathrm{cut}}\leqslant 0.41$ (middle row), and $0.7\leqslant \omega_{\bm{K}}/\omega_{\mathrm{cut}}\leqslant 0.71$ (bottom row). 
The Thomson distributions are presented as the mirror-reflected curves. They were obtained after applying the frequency scaling \eqref{fre5}
but without multiplying them by the factor $\gamma(\omega_{\bm{K}},\Omega_{\bm{K}})$, in order to show their absolute values. In the left column, 
we show the Compton distributions for both no-spin-flipping (solid blue) and spin-flipping (dashed magenta) processes, and for the emitted radiation polarized in the scattering plane. 
As one can see, the spin-flipping processes marginally contribute to the total emitted energy. It is interesting to note that for all these intervals 
the Thomson and the no-spin-flipping Compton distributions are similar in the sense discussed above, although they are calculated for frequency domains 
that are very much separated from each other. For instance, in the bottom left panel the Compton and Thomson processes are calculated for 
$0.35\leqslant \omega_{\bm{K}}/m_{\mathrm{e}}c^2\leqslant 0.355$ and $1.17\leqslant \omega_{\bm{K}}^{\mathrm{Th}}/m_{\mathrm{e}}c^2\leqslant 1.22$, respectively. 
This proves the validity of the classical theory (up to the frequency scaling) for frequencies $\omega_{\bm{K}}$ not significantly smaller than $\omega_{\mathrm{cut}}$. 

In the right column of Fig.~\ref{ctcheck3a}, we present the Compton distribution for the emitted radiation of polarization perpendicular 
to the scattering plane. While for small frequencies (top panel), the no-spin-flipping process dominates, thus with increasing the frequency range
of emitted radiation the spin-flipping process starts to play a role. In fact, there are some frequency domains for which the process 
that does not conserve the electron spin occurs with by far more significant probability than a process that does conserve the electron spin 
(see, also Ref.~\cite{KKpol} and the discussion in Sec. VB). This becomes even more clear for frequencies closer to the threshold value $\omega_{\mathrm{cut}}$.

\begin{figure}
\includegraphics[width=7.0cm]{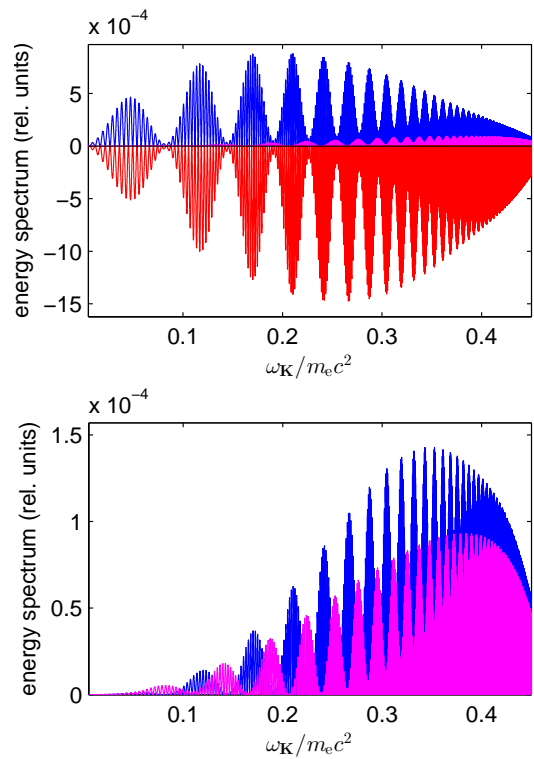}%
\caption{(Color online) Energy spectra for the Compton scattering (the solid blue line is for the no-spin-flipping process, $\lambda_{\mathrm{i}}\lambda_{\mathrm{f}}=1$, 
the solid magenta (light gray) line is for the spin-flipping process, $\lambda_{\mathrm{i}}\lambda_{\mathrm{f}}=-1$), Eq.~\eqref{com2}, and for the Thomson scattering (red line, reflected with respect to the horizontal black line), 
Eq.~\eqref{thom5}. The presented results are for the laser pulse propagating in the $z$-direction with a linear polarization vector along the $x$-axis.
The remaining parameters are: $\mu=10$, $N_{\mathrm{osc}}=2$, and $\omega_{\mathrm{L}}=0.3 m_{\mathrm{e}}c^2$. The direction of scattered radiation 
is given by the polar and azimuthal angles $\theta_{\bm{K}}=0.99\pi$ and $\varphi_{\bm{K}}=0$. This parameters are in the reference frame of incident electrons. 
In the upper panel, the energy spectra are presented for radiation emitted with a linear polarization in the scattering plane. In the lower panel,
the energy spectra of Compton radiation polarized perpendicularly to the scattering plane are displayed; note that in this case the Thomson theory gives 0. 
The Thomson spectrum is calculated for the frequency $\omega_{\bm{K}}^{\mathrm{Th}}$ and then the frequency is transformed to the Compton frequency 
$\omega_{\bm{K}}$ by applying the scaling law \eqref{fre5}. For these particular parameters and for the chosen reference frame, $\omega_{\mathrm{cut}}\approx m_{\mathrm{e}}c^2/2$.
\label{ctcheck3sb}}
\end{figure} 

When comparing the corresponding panels in different columns of Fig.~\ref{ctcheck3a}, one can conclude that the emission of Compton photons polarized
perpendicularly to the scattering plane is suppressed as compared to the emission of photons polarized in that plane. However, we showed in Ref.~\cite{KKpol}
that this is not always the case. This appears to be a purely quantum effect, as classically there is no emission of perpendicularly polarized 
emitted radiation (see, the right column of Fig.~\ref{ctcheck3a}).

\subsection{Short laser pulses}

In this section, we consider very short laser pulses with $N_{\mathrm{osc}}=2$. In this case, the four-momentum conservation condition, Eq.~\eqref{fre1}, is 
rather vaguely satisfied for an integer $N_{\mathrm{eff}}$, due to the time-energy uncertainty relation. In other words, contrary to long pulses, the final electron 
momenta $p_{\mathrm{f}}$ in Eq.~(28) for which $N_{\mathrm{eff}}$ is not an integer, significantly contribute to the sum in Eq.~(15). 
Nevertheless, we observe a very good agreement between the quantum and classical theories. In Fig.~\ref{ctcheck3sb}, we compare the Compton and Thomson distributions for the same geometry and the same 
laser field parameters as in Fig.~\ref{ctcheck3a}, except that the number of field oscillations within the pulse is small. In the upper panel, the polarization 
of emitted radiation is in the scattering plane. For very short laser pulses, the spin-flipping processes play a more significant role. Moreover, up to a multiplicative 
factor we find very good agreement between the no-spin-flipping Compton scattering and the Thomson one for frequencies close to the cut-off value, 
$\omega_{\mathrm{cut}}$. On the other hand, for the polarization perpendicular to the scattering plane, the spin-flipping process dominates for some frequency 
domains over the no-spin-flipping one (the lower panel in Fig.~\ref{ctcheck3sb}). Note also that independent of the polarization of emitted radiation, for frequencies 
close to the cut-off frequency, both the spin-flipping and the no-spin-flipping processes occur with comparable probabilities. 

\begin{figure}
\includegraphics[width=7.0cm]{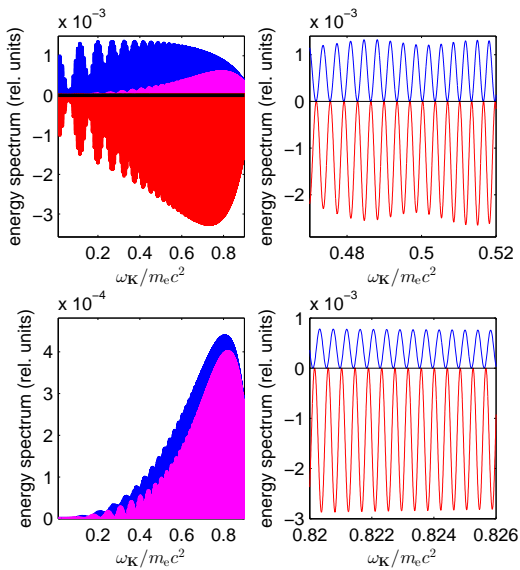}%
\caption{(Color online) The left column represents the same as in Fig. \ref{ctcheck3sb}, but with $\theta_{\bm{K}}=0.5\pi$, for which $\omega_{\mathrm{cut}}= m_{\mathrm{e}}c^2$. 
The right column shows two enlarged parts of the upper-left frame, but only for the no-spin-flipping processes. We observe very good agreement between 
the Compton and Thomson results (up to the multiplicative factor) even for $\omega_{\bm{K}}/\omega_{\mathrm{cut}}$ close to 1.
\label{ctcheck31sd}}
\end{figure}

These general features are confirmed for scattering at a smaller polar angle, $\theta_{\bm{K}}=0.5\pi$, which is equivalent to the larger cut-off 
frequency ($\omega_{\mathrm{cut}}=m_{\mathrm{e}}c^2$), as presented in Fig.~\ref{ctcheck31sd}. In this case, the spin-flipping and the no-spin-flipping 
processes become almost equal for large frequencies (cf., the left column). Moreover, similarities between the quantum and the classical treatments 
survive even for frequencies from the domain of $0.82\leqslant \omega_{\bm{K}}/m_{\mathrm{e}}c^2\leqslant 0.826$ (the lower panel in the right column), 
although we start to observe here a tiny blue-shift of the classical distribution after applying the frequency transformation~\eqref{fre5}.

\section{Prior analysis of the scaling law for finite incident laser pulses}

The frequency scaling of emitted radiation for Compton and Thomson processes induced by finite laser pulses was introduced 
in Ref.~\cite{Seipt}. This transformation was defined for finite but sufficiently long laser pulses. Here, we relate our results
to the work of Seipt and K\"ampfer~\cite{Seipt}.

Seipt and K\"ampfer have started their discussion of the frequency scaling law by introducing the dimensionless and relativistically 
invariant parameter $y_{\ell}$ (Eq. (2) in Ref.~\cite{Seipt}), which in our notation equals
\begin{equation}
y_{\ell}=2\ell\omega_{\mathrm{L}}\frac{n\cdot p_{\mathrm{i}}}{m_{\mathrm{e}}^2c^3} .
\label{SK1}
\end{equation}
For a monochromatic plane wave field, $\ell$ is interpreted as the number of laser photons absorbed during the Compton scattering 
(at least for not too intense laser fields). As it has been remarked by the authors (see, comment after Eq. (50) of Ref.~\cite{Seipt}), the parameter $\ell$ is 
inappropriate for finite pulses because the energy distribution of emitted photons becomes a continuous function of the frequency 
$\omega_{\bm{K}}$. Moreover, for sufficiently intense laser pulses, measured by the parameter $a_0$ (which is related to our $\mu$ 
and, in fact, equals $\mu$ for a monochromatic plane wave field), some parts of the energy distribution are not conclusively labeled by $\ell$ 
(see, e.g., Fig.~8 in \cite{Seipt}). This is the reason why the parameter $\ell$ for finite and short laser pulses does not have 
physical meaning and should be entirely eliminated from the formulation and discussion of the scaling law, as it has been done in our 
analysis. It is still easy to establish the connection of the parameter $\ell$ with our $N_{\mathrm{eff}}$; namely,
\begin{equation}
\ell=N_{\mathrm{eff}}/N_{\mathrm{osc}}.
\label{SK2}
\end{equation}
The point is that $N_{\mathrm{eff}}\omega=\ell\omega_{\mathrm{L}}$ corresponds to the most probable energy absorbed from the laser pulse in order 
to generate the Compton photon of four-momentum $K$. Let us stress, however, that this relation has the probabilistic interpretation and it does not mean 
that for a finite laser pulse the four-momentum conservation equation \eqref{fre1} is fulfilled; it only becomes more \textit{probable} as the duration 
of the pulse increases.

By analyzing the integrated distributions for both processes, when driven by not very intense laser pulses, 
it has been found in~\cite{Seipt} that both classical and quantum approaches give the same values for $y_1 \lesssim 10^{-2}$ (cf. Fig.~3 in \cite{Seipt}). 
For larger values of $y_1$, the Thomson scattering signal becomes much larger than the signal of Compton scattering. Since small values of $y_1$ 
correspond to low frequencies, $\omega_{\bm{K}}\ll\omega_{\mathrm{cut}}$, therefore our results for differential distributions 
are in full agreement with this statement. Note, that the angle-integrated cross sections have been calculated in~\cite{Seipt} 
for a small intensity of the laser pulse (i.e., $a_0\ll 1$), whereas for higher intensities (with $a_0 \leqslant 2$) only results 
for the fully differential distributions have been presented. In our studies so far, we have also presented only differential distributions;
except that we have considered more intense laser pulses. The point is that subtle peak structures observed in the fully differential 
distributions are washed out in the angle-integrated distributions; hence, a detailed theoretical comparison of quantum and classical approaches is not possible.

\begin{figure}
\includegraphics[width=7.0cm]{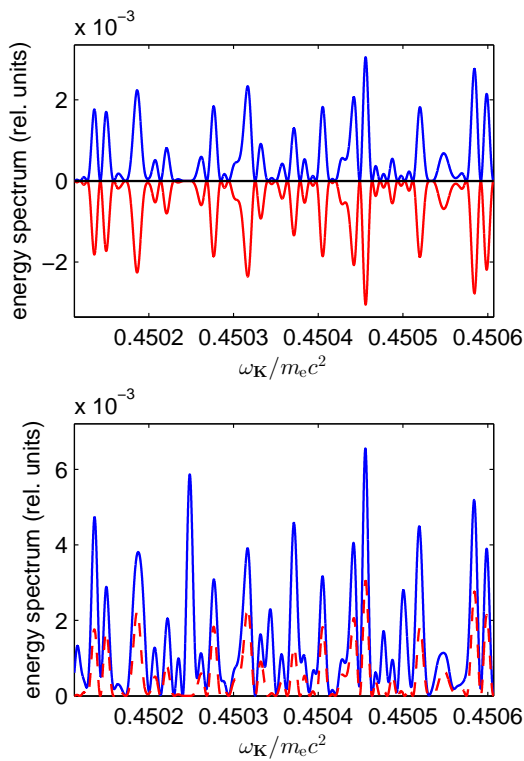}%
\caption{(Color online) The same as in Figs.~\ref{ctcheck3ff} and~\ref{ctcheck3a}, but for $\omega_{\bm{K}}$ very close to the cut-off 
value $\omega_{\mathrm{cut}}$. In the upper panel, the Compton (for the spin-conserved process and with the Compton photon linearly 
polarized in the scattering plane) and Thomson (mirror-reflected) energy spectra are compared such that the Thomson distribution is 
normalized to the maximum value of the Compton one. Although in this frequency domain both distributions exhibit rather irregular behavior, 
after frequency transformation and normalization there is the perfect agreement between quantum spin-conserved and classical theories. 
In the lower panel, the Compton distribution from the upper frame (dashed line) is compared to the total energy distribution for the 
Compton process (solid line), when summed over all final spin and polarization degrees of freedom, and averaged over the initial spins. 
In this frequency domain all spin and polarization degrees of freedom contribute significantly to the total distribution. 
\label{compare2}}
\end{figure}

Consider the case of a long laser pulse with $N_{\mathrm{osc}}=16$, for which the differential distributions are shown in Fig.~\ref{ctcheck3a}. 
For these laser field parameters, $y_1=2\omega_{\mathrm{L}}/m_{\mathrm{e}}c^2=0.6$. In the top row of Fig.~\ref{ctcheck3a},
$N_{\mathrm{eff}}$ changes from 58.7 to 65.3, which means that $\ell\approx 4$. Although $y_{\ell}$ is comparable to 1, 
we observe perfect agreement between the spin-conserved Compton and the frequency-scaled Thomson distributions, which also holds for their absolute values. 
With increasing $\ell$, the absolute values of the frequency-scaled Thomson distribution starts to dominate over the Compton distribution.
Still, the positions of extrema and the structure of these distributions stay the same (for the middle row in Fig.~\ref{ctcheck3a} we have 
$\ell\approx 22$, and for the bottom row $\ell\approx 80$). Let us further investigate a more extreme case presented in the upper panel 
of Fig.~\ref{compare2} for frequencies of generated Compton photons very close to the cut-off value $\omega_{\mathrm{cut}}$. In this frequency domain, 
$\ell\approx 300$. This is the case of the 'overlapping' harmonics (specified by the condition (50) in~\cite{Seipt}), or the 'erratic' 
(irregular) part of the Compton distribution (as discussed in Sec. IV.D in \cite{Seipt}). Again, we observe perfect agreement 
(up to a normalization) between the frequency-scaled Thomson and no-spin-flipping Compton distributions. We conclude that once the spin and polarizations effects are
accounted for, the scaling law is applicable in the region where the spectral densities show the erratic behavior. To confirm this statement,
in the lower panel of Fig.~\ref{compare2}, the spin-conserved Compton distribution (dashed line) is compared to the total 
Compton distribution (solid line); the latter being summed over the final electron spin and photon polarization degrees of freedom and averaged 
over the initial electron spins. Here, contrary to the low-frequency case where the spin-conserved process dominates, the polarization and 
spin effects for high-frequency part of the Compton distribution cannot be considered as trivial. 
Let us also remark that in Ref. \cite{Seipt} it has been suggested that for the erratic part of the spectrum, 
where the individual harmonics overlap, the classical radiation reaction force presumably should be accounted for in calculations  
as it introduces an extra electron recoil in Thomson scattering. Such a statement could be valid, but our analysis also shows that 
the erratic behavior in the emitted spectrum appears when the spin-flipping process starts to be important. One can anticipate that
interferences between probability amplitudes with different electron spins can result in the erratic behavior, observed in Ref.~\cite{Seipt}.

It is commonly accepted that strong-field QED is the proper theoretical scheme for the analysis of high energy photons generated by 
the interaction of electrons with strong laser pulses. It is also understood that the classical theory can be only considered 
as its approximation. Our investigations show that for some parts of the spectrum the spin-flipping process occurs with a significant 
probability distribution. Therefore, one can assume that for spin-polarized electrons it is experimentally feasible to detect 
the spin-flipping Compton process. Our analysis can suggest the most suitable parameters for such a detection.

For shorter laser pulses (with smaller $N_{\mathrm{osc}}$), the spin and polarization effects become even more important. This is observed in 
Figs.~\ref{ctcheck3sb} and~\ref{ctcheck31sd} for $N_{\mathrm{osc}}=2$. Nevertheless, for spin-conserved Compton and frequency-scaled Thomson 
processes we still observe structural similarity (i.e., the coincidence in the positions and relative values of peaks). This supports 
the postulate formulated in Ref.~\cite{Seipt} that the scaling law may be applied for arbitrary laser beams, including short laser beams. While this
is proven in our paper for the first time, let us mention an important aspect of our formulation. As we have emphasized in our previous 
publications~\cite{KKcompton,KKbw}, for finite laser pulses, the laser-field-dressing of the initial and final electron momenta 
differs from the dressing induced by a monochromatic plane wave. Namely, apart from the terms proportional to the time-averaged 
shape functions squared, $\langle f_i^2\rangle$ (which lead to the effective electron mass in the field), there are also terms proportional 
to $\langle f_i\rangle$ and to the polarization vectors of the pulse [Eqs.~\eqref{com0} and~\eqref{fre2a}]. These terms are responsible for 
angular asymmetries in various strong-field QED processes (see, e.g., Refs.~\cite{KKcompton,KKbw}). In addition, these terms lead 
to a redefinition of the Thomson (classical) frequency for short laser pulses, Eq.~\eqref{fre3}, which now becomes the laser-field-polarization dependent. 
Such a correction of the classical frequency, which is \textit{vital} for short driving pulses, is not accounted for in the Seipt-K\"ampfer definition 
of the classical frequency (Eq. (55) in~\cite{Seipt} with the definition of the four-momentum $q$ in the text), used further in their 
formulation of the scaling law (Eqs. (56) and (57) in \cite{Seipt}).

\section{Angular distributions}

In the prior analysis of the scaling law \cite{Heinzl,HSK,Seipt} only the frequency distribution for the Compton and Thomson processes 
has been studied. The aim of the remaining part of this paper is to extend the validity of the frequency scaling law discussed above 
and to investigate the angular distributions as well as the temporal power distribution of emitted radiation.

\subsection{Polar-angle distribution}

In polar-angle distributions plotted in this section, we fix the frequency $\omega_{\bm{K}}$ and the azimuthal angle of emitted
radiation $\varphi_{\bm{K}}$, whereas we change its polar angle $\theta_{\bm{K}}$. 
When comparing the Compton and the frequency-scaled Thomson spectra we have to remember that the cut-off frequency 
$\omega_{\mathrm{cut}}$ depends on the direction of emitted radiation, which introduces an extra angular 
dependence into the scaled Thomson amplitude. In addition, we have found it more convenient to plot the spectra 
as the function of angles $(\Theta_{\bm{K}},\Phi_{\bm{K}})$, $0\leqslant \Theta_{\bm{K}}<2\pi$ and $0\leqslant \Phi_{\bm{K}}<\pi$, such that
\begin{equation}
(\theta_{\bm{K}},\varphi_{\bm{K}})=\begin{cases}
(\Theta_{\bm{K}},\Phi_{\bm{K}}), & \mathrm{for}\quad 0\leqslant\Theta_{\bm{K}}\leqslant\pi,
\cr
(2\pi-\Theta_{\bm{K}},\Phi_{\bm{K}}+\pi), & \mathrm{for}\quad \pi<\Theta_{\bm{K}}<2\pi ,
\end{cases}
\label{newangles}
\end{equation}
which accounts for the continuity of the distributions at the south pole.

\begin{figure}
\includegraphics[width=7.0cm]{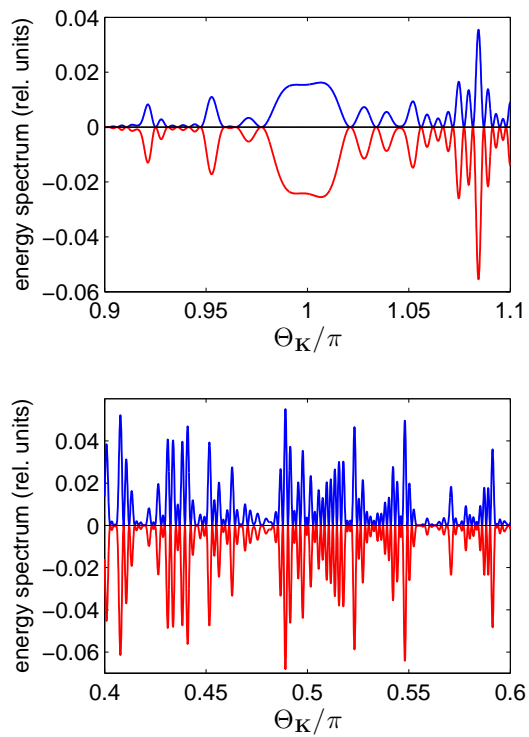}%
\caption{(Color online) The same as in Fig.~\ref{ctcheck3a}, but only for the no-spin-flipping process ($\lambda_{\mathrm{i}}\lambda_{\mathrm{f}}=1$), fixed frequency of generated radiation, $\omega_{\bm{K}}=0.2024m_{\mathrm{e}}c^2$, $\Phi_{\bm{K}}=0$, and for two domains of the emission angle $\Theta_{\bm{K}}$. In both cases the polar-angle distributions for the Compton (blue line) and Thomson (mirrored red line) scattering show the structural similarity as for the frequency distributions.
\label{theta16osc3and2d20140809}}
\end{figure}

In Fig.~\ref{theta16osc3and2d20140809}, we compare the spin-conserved Compton polar-angle distribution 
with the respective frequency-scaled Thomson distribution for the given frequency 
$\omega_{\bm{K}}=0.2024m_{\mathrm{e}}c^2$, $\Phi_{\bm{K}}=0$, and for a long driving pulse 
with $N_{\mathrm{osc}}=16$. We observe, similarly to the frequency distributions presented in Fig.~\ref{ctcheck3a}, 
the perfect structural agreement (i.e., maxima and zeros of both distributions appear for the same polar angle 
$\Theta_{\bm{K}}$) between the quantum and frequency-scaled classical theories. Exactly the same agreement is observed
for very short pulses, with $N_{\mathrm{osc}}=2$. Such an agreement is not achievable if the frequency scaling law 
proposed in Refs.~\cite{Heinzl,HSK,Seipt} is applied, as these works are missing the term with $\langle f_1\rangle$ 
in the momentum dressing, Eq.~\eqref{com0}. Inspection of Fig.~\ref{theta16osc3and2d20140809} also shows that the 
structural similarity between the quantum and frequency-scaled classical theories appears for the so-called 
\textit{regular} part of the distributions (i.e., for emission angles close to the south pole, $\theta_{\bm{K}}\approx \pi$) 
as well as for the \textit{irregular} part (i.e., for emission angles close to the equatorial, $\theta_{\bm{K}}\approx \pi/2$).
Even though we observe there rapid changes of the intensity of generated radiation with densely distributed maxima.

A similar agreement between the Compton and the frequency-scaled Thomson distributions exist also for non-zero 
$\Phi_{\bm{K}}$, provided that polarization vectors of emitted radiation are suitably chosen. Since the same concerns 
azimuthal-angle distributions, this problem will be discussed in the following section.

\subsection{Azimuthal-angle distribution}

As we have already stressed, polarization properties of the emitted radiation play the crucial role in the analysis of 
the frequency scaling. Therefore, let us first define the convention of how the polarization vectors are introduced 
in our numerical analysis. The two linear polarizations, $\bm{\varepsilon}_{\bm{K},1}$ and $\bm{\varepsilon}_{\bm{K},2}$, 
are fixed such that for radiation generated in the direction $\bm{n}_{\bm{K}}$ (determined by the polar and azimuthal angles, 
$\theta_{\bm{K}}$ and $\varphi_{\bm{K}}$) the three vectors (see, Appendix A in \cite{KTK2014b}),
\begin{align}
\bm{\varepsilon}_{\bm{K},1}=&\begin{pmatrix}\cos\theta_{\bm{K}}\cos\varphi_{\bm{K}} \cr \cos\theta_{\bm{K}}\sin\varphi_{\bm{K}} \cr -\sin\theta_{\bm{K}} \end{pmatrix} ,\,
\bm{\varepsilon}_{\bm{K},2}=\begin{pmatrix}-\sin\varphi_{\bm{K}} \cr \cos\varphi_{\bm{K}} \cr 0 \end{pmatrix} , \nonumber \\
\bm{n}_{\bm{K}}=&\begin{pmatrix}\sin\theta_{\bm{K}}\cos\varphi_{\bm{K}} \cr \sin\theta_{\bm{K}}\sin\varphi_{\bm{K}} \cr  \cos\theta_{\bm{K}} \end{pmatrix} ,
\label{app1}
\end{align}
create the right-hand-side system of orthogonal unit vectors, 
\begin{equation}
\bm{\varepsilon}_{\bm{K},1}\times \bm{\varepsilon}_{\bm{K},2}=\bm{n}_{\bm{K}}.
\label{aa2}
\end{equation}
Since the Thomson and Compton processes are relativistically invariant, we can choose the Lorentz reference frame such 
that the laser beam and the electron counterpropagate. Next, we can orient the coordinate system such that 
\begin{equation}
\bm{\varepsilon}_1=\bm{e}_x, \, \bm{\varepsilon}_2=\bm{e}_y, \, \bm{n}=\bm{e}_z.
\label{aa3}
\end{equation}
In this paper we consider a linearly polarized laser pulse for which the second shape function vanishes, $f_2(k\cdot x)=0$. 
This allows us to define the laser pulse-plane, spanned by vectors $\bm{\varepsilon}_1$ and $\bm{n}$, in which the classical 
motion of electrons takes place. This means that the vector $\bm{\Upsilon}(\phi)$ [Eq.~\eqref{thom11}] is coplanar with this plane. 
Hence, these parts of the polarization vectors of emitted radiation that are perpendicular to the laser pulse-plane do not 
contribute to the Thomson amplitude, which is not the case for the Compton amplitude. In order to compare reasonably 
predictions of the classical and quantum theories for an arbitrary direction of emission we have to choose a different convention 
for the polarization vectors of emitted radiation. This can be done along the line suggested in Ref.~\cite{KTK2014b} (see, Appendix A). 
Namely, instead of $\bm{\varepsilon}_{\bm{K},1}$ and $\bm{\varepsilon}_{\bm{K},2}$ we choose the following unit vectors:
\begin{equation}
\bm{\varepsilon}_{\bm{K},\|}=\frac{(\bm{\varepsilon}_{\bm{K},2}\cdot\bm{\varepsilon}_2)\bm{\varepsilon}_{\bm{K},1}
-(\bm{\varepsilon}_{\bm{K},1}\cdot\bm{\varepsilon}_2)\bm{\varepsilon}_{\bm{K},2}}
{\sqrt{(\bm{\varepsilon}_{\bm{K},2}\cdot\bm{\varepsilon}_2)^2+(\bm{\varepsilon}_{\bm{K},1}\cdot\bm{\varepsilon}_2)^2}},
\label{aa4}
\end{equation}
\begin{equation}
\bm{\varepsilon}_{\bm{K},\bot}=\frac{(\bm{\varepsilon}_{\bm{K},1}\cdot\bm{\varepsilon}_2)\bm{\varepsilon}_{\bm{K},1}
+(\bm{\varepsilon}_{\bm{K},2}\cdot\bm{\varepsilon}_2)\bm{\varepsilon}_{\bm{K},2}}
{\sqrt{(\bm{\varepsilon}_{\bm{K},2}\cdot\bm{\varepsilon}_2)^2+(\bm{\varepsilon}_{\bm{K},1}\cdot\bm{\varepsilon}_2)^2}},
\label{aa5}
\end{equation}
which also form the right-hand-side system of orthogonal unit vectors, 
\begin{equation}
\bm{\varepsilon}_{\bm{K},\|}\times \bm{\varepsilon}_{\bm{K},\bot}=\bm{n}_{\bm{K}}.
\label{aa6}
\end{equation}
With these new polarization vectors we can define the Thomson amplitudes,
\begin{equation}
\mathcal{A}_{\mathrm{Th},\|}=
\frac{(\bm{\varepsilon}_{\bm{K},2}\cdot\bm{\varepsilon}_2)\mathcal{A}_{\mathrm{Th},1}
-(\bm{\varepsilon}_{\bm{K},1}\cdot\bm{\varepsilon}_2)\mathcal{A}_{\mathrm{Th},2}}
{\sqrt{(\bm{\varepsilon}_{\bm{K},2}\cdot\bm{\varepsilon}_2)^2+(\bm{\varepsilon}_{\bm{K},1}\cdot\bm{\varepsilon}_2)^2}},
\label{aa7}
\end{equation}
\begin{equation}
\mathcal{A}_{\mathrm{Th},\bot}=
\frac{(\bm{\varepsilon}_{\bm{K},1}\cdot\bm{\varepsilon}_2)\mathcal{A}_{\mathrm{Th},1}
+(\bm{\varepsilon}_{\bm{K},2}\cdot\bm{\varepsilon}_2)\mathcal{A}_{\mathrm{Th},2}}
{\sqrt{(\bm{\varepsilon}_{\bm{K},2}\cdot\bm{\varepsilon}_2)^2+(\bm{\varepsilon}_{\bm{K},1}\cdot\bm{\varepsilon}_2)^2}},
\label{aa8}
\end{equation}
and similarly the spin-dependent Compton amplitudes. In analogy to Eqs.~\eqref{com2} and \eqref{thom6}, we introduce 
the energy distributions for the emitted radiation for these two particular polarization vectors. 

\begin{figure}
\includegraphics[width=7.0cm]{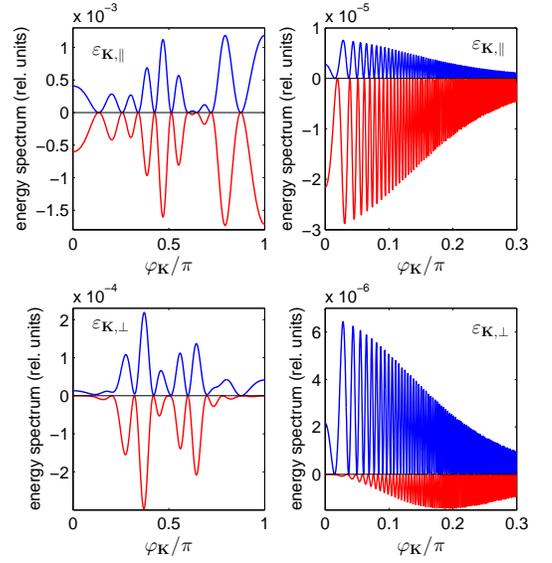}%
\caption{(Color online) Azimuthal-angle distribution of radiation energy generated by the no-spin-flipping Compton process 
(blue line) and by the frequency-scaled Thomson scattering (mirrored red line) for the same laser pulse parameters as in 
Figs.~\ref{ctcheck3sb} and \ref{ctcheck31sd}. The upper row presents distributions of radiation linearly polarized 
in the laser pulse-plane [cf. Eq.~\eqref{aa4}], and the lower row for polarization \eqref{aa5}. In the left column 
we show the results for $\omega_{\bm{K}}=0.2024m_{\mathrm{e}}c^2$, and in the right one for $\omega_{\bm{K}}=0.6m_{\mathrm{e}}c^2$. 
For all cases the polar angle $\theta_{\bm{K}}=0.7\pi$. These distributions satisfy the symmetry $\varphi_{\bm{K}}\rightarrow 2\pi-\varphi_{\bm{K}}$.
\label{phi2and3}}
\end{figure}

In Fig.~\ref{phi2and3}, we compare the azimuthal-angle distributions for the spin-conserved Compton and frequency-scaled 
Thomson processes for the short laser pulse, $N_{\mathrm{osc}}=2$, and for two linear polarizations of emitted radiation 
defined by the vectors \eqref{aa4} and \eqref{aa5}. As anticipated, we observe a very good agreement between the results for polarization parallel 
to the laser pulse-plane, and a disagreement for the second polarization vector. The point being that, in general, it has 
a non-vanishing component perpendicular to the laser pulse-plane, not accounted for by the classical theory. Indeed, 
a closer look at the lower row of this figure (for the polarization $\bm{\varepsilon}_{\bm{K},\bot}$) shows that, although 
positions of maxima and zeros are nearly the same, the coarse-grained quantum and classical distributions are different.
This is particularly well-manifested for larger frequencies $\omega_{\bm{K}}$. Let us also note that, for frequencies of 
generated radiation that are closer to the cut-off frequency for Compton scattering, we start 
observing a tiny shift for the classical azimuthal-angle distribution (cf. the upper right frame in Fig.~\ref{phi2and3}).
This is similar to the small frequency blue shift detected for the frequency distribution (cf. the lower right frame in Fig.~\ref{ctcheck31sd}). 
It remains an open question whether such tiny discrepancies between the quantum and frequency-scaled classical distributions 
can be corrected by the classical radiation reaction \cite{LL2,Nakhleh2012,Goetz1,Goetz2}, which introduces 
the recoil of electrons during their interaction with the laser pulse.

\section{Total energy of generated radiation}

In this section, we present the results for the total energy of radiation generated from Compton and Thomson processes. 
In the case of Compton scattering, we have to perform the three-dimensional integral which we write as
\begin{equation}
E_{\mathrm{C}}=\int\limits_{0}^{2\pi}\mathrm{d}\varphi_{\bm{K}}\int\limits_{-1}^{1}\mathrm{d}\cos\theta_{\bm{K}}
\int\limits_{0}^{\omega_{\mathrm{cut}}}\mathrm{d}\omega_{\bm{K}}\, F_{\mathrm{C}}(\omega_{\bm{K}},\theta_{\bm{K}},\varphi_{\bm{K}}).
\label{tot1}
\end{equation}
Note that, in general, the cut-off frequency depends on angles, 
$\omega_{\mathrm{cut}}=\omega_{\mathrm{cut}}(\theta_{\bm{K}},\varphi_{\bm{K}})$ [although, for the head-on collision 
considered in this paper it is $\varphi_{\bm{K}}$-independent], and [cf. Eq.~\eqref{com1}]
\begin{equation}
F_{\mathrm{C}}(\omega_{\bm{K}},\theta_{\bm{K}},\varphi_{\bm{K}})=\frac{\mathrm{d}^3E_{\mathrm{C}}}{\mathrm{d}\omega_{\bm{K}}\mathrm{d}^2\Omega_{\bm{K}}}.
\label{tot2}
\end{equation}
Changing the parameters ($0\leqslant\xi_i\leqslant 1$, $i=1,2,3$),
\begin{equation}
\varphi_{\bm{K}}=2\pi\xi_1, \, \cos\theta_{\bm{K}}=2\xi_2-1, \, \omega_{\bm{K}}=\omega_{\mathrm{cut}}\xi_3,
\label{tot3}
\end{equation}
we arrive at the three-dimensional integral over the unit cube,
\begin{equation}
E_{\mathrm{C}}=\int\limits_{0}^{1} \mathrm{d}\xi_1\mathrm{d}\xi_2\mathrm{d}\xi_3 \, \tilde{F}_{\mathrm{C}}(\xi_1,\xi_2,\xi_3),
\label{tot4}
\end{equation}
where
\begin{equation}
\tilde{F}_{\mathrm{C}}(\xi_1,\xi_2,\xi_3)=4\pi\omega_{\mathrm{cut}}(\theta_{\bm{K}},\varphi_{\bm{K}})F_{\mathrm{C}}(\omega_{\bm{K}},\theta_{\bm{K}},\varphi_{\bm{K}}).
\label{tot5}
\end{equation}

\begin{figure}
\includegraphics[width=7.0cm]{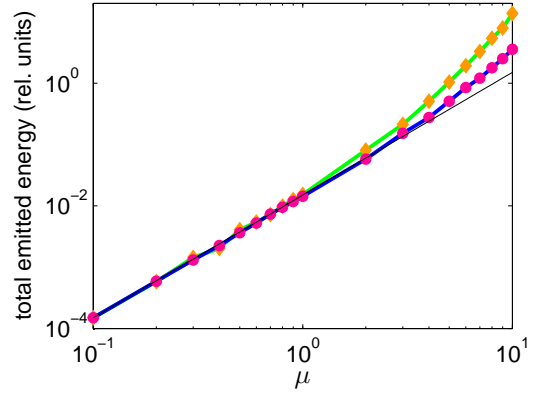}%
\caption{(Color online) Total energy (in relativistic units) in the electron's reference frame for linearly polarized laser pulse defined by Eq.~\eqref{las8}, $\omega_{\mathrm{L}}=0.3m_{\mathrm{e}}c^2$ and $N_{\mathrm{osc}}=16$. In the log-log plot we present the total energy calculated for the Compton process (dark red circles) and for the Thomson process (light brown diamonds). The continuous blue (dark) and green (light) lines are to guide the eye. The thin black straight line represents the fitting curve defined by Eq.~\eqref{fitting}. 
\label{total16osc}}
\end{figure}

For the Thomson scattering the only difference is that the integration over $\omega_{\bm{K}}$ in Eq.~\eqref{tot1} extends 
to infinity. Applying the frequency scaling, Eq.~\eqref{fre6}, we obtain in a very similar way the expression for the total energy of radiation generated by the classical process,
\begin{equation}
E_{\mathrm{Th}}=\int\limits_{0}^{1} \mathrm{d}\xi_1\mathrm{d}\xi_2\mathrm{d}\xi_3 \, \tilde{F}_{\mathrm{Th}}(\xi_1,\xi_2,\xi_3),
\label{tot6}
\end{equation}
where
\begin{equation}
\tilde{F}_{\mathrm{Th}}(\xi_1,\xi_2,\xi_3)=
\frac{4\pi\omega_{\mathrm{cut}}(\theta_{\bm{K}},\varphi_{\bm{K}})}{\bigl(1-\frac{\omega_{\bm{K}}}{\omega_{\mathrm{cut}}(\theta_{\bm{K}},\varphi_{\bm{K}})}\bigr)^2}
F_{\mathrm{Th}}(\omega_{\bm{K}},\theta_{\bm{K}},\varphi_{\bm{K}})
\label{tot7}
\end{equation}
and
\begin{equation}
F_{\mathrm{Th}}(\omega_{\bm{K}},\theta_{\bm{K}},\varphi_{\bm{K}})=\frac{\mathrm{d}^3E_{\mathrm{Th}}}{\mathrm{d}\omega_{\bm{K}}^{\mathrm{Th}}\mathrm{d}^2\Omega_{\bm{K}}} \bigg|_{\omega_{\bm{K}}^{\mathrm{Th}}=\frac{\omega_{\bm{K}}}{1-\frac{\omega_{\bm{K}}}{\omega_{\mathrm{cut}}(\theta_{\bm{K}},\varphi_{\bm{K}}) } }}  .
\label{tot8}
\end{equation}

For large laser field intensities the integrands in Eqs. \eqref{tot4} and \eqref{tot6} are rapidly changing functions of their arguments, 
which makes the standard multi-dimensional integration algorithms (usually based on the Gauss-type methods) hardly applicable. 
However, similar to the Bethe-Heitler process \cite{KMK2013,KKE2006,KKrec}, we have found that the Monte Carlo algorithm is sufficiently 
fast convergent, with the estimated error not larger than a few percents for $10^6$ sample points.

\begin{figure}
\includegraphics[width=7.0cm]{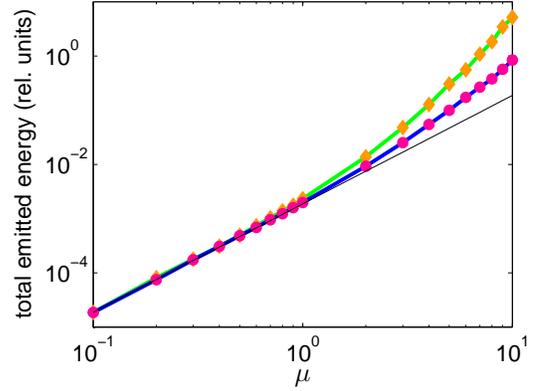}%
\caption{(Color online) The same as in Fig.~\ref{total16osc} but for $N_{\mathrm{osc}}=2$.
\label{total2osc}}
\end{figure}

In Figs.~\ref{total16osc} and \ref{total2osc}, we present in the log-log plots the estimated values for the total energy of generated 
radiation in the electron's reference frame as functions of the relativistically invariant parameter $\mu$ for the long and short laser 
pulses (i.e., for $N_{\mathrm{osc}}=16$ and $N_{\mathrm{osc}}=2$, respectively). One can see that for the intensity parameter $\mu$ not 
larger that 1 the markers lie on the straight line. We have found that in this region the dependence on $\mu$ of the total emitted 
energy for Compton and Thomson scattering fits the curve 
\begin{equation}
E(\mu)=E_0 N_{\mathrm{osc}} \mu^2,
\label{fitting}
\end{equation}
with $E_0\approx 0.94\times 10^{-3}m_{\mathrm{e}}c^2$. This parameter is a universal quantity in the sense that for the considered in 
this paper shape of the laser pulse, Eq.~\eqref{las8}, it is independent of the number of oscillations $N_{\mathrm{osc}}$ 
(we have checked this also for $N_{\mathrm{osc}}=8$). For $\mu$ larger than 1 the total emitted energy starts increasing with $\mu^2$ nonlinearly.

We learn from Figs.~\ref{total16osc} and \ref{total2osc} that for laser field intensities such that $\mu\lesssim 1$ (for the Ti-sapphire laser $\mu=1$ 
corresponds to the intensity of the order of $10^{18}\mathrm{W/cm}^2$) the quantum and classical approaches give nearly the same results. 
This can be expected as for such intensities only radiation of frequencies $\omega_{\bm{K}}$ much smaller than the cut-off one is generated 
with a significant probability. Hence, Compton and Thomson (even without the frequency scaling) theories predict nearly identical differential 
distributions. Discrepancies start to be visible for larger intensities. Specifically, for $\mu\approx 10$ (i.e., for intensities of the order 
of $10^{20}\mathrm{W/cm}^2$ for the Ti-sapphire laser) the classical predictions exceed the quantum ones even by one order of magnitude. 
One can anticipate that for still larger intensities ($\mu\gtrsim 100$, as expected for instance in the ELI \cite{eli} or XCELS \cite{exel} projects) 
the differences between results based on the classical and quantum approaches can be even larger. For such intensities, the strong-field QED 
analysis of fundamental processes is very much demanding, or sometimes even impossible, to be carried out. For this reason, the approach based on 
classical electrodynamics is mostly applied (for relevant review articles, see, e.g., Refs. \cite{Pukhov2003,TG2009}). For instance, the problem 
of generation of zeptosecond (or even yoctosecond) pulses is currently vigorously studied (see, e.g., \cite{Galkin2009,Chung2009,Liu2012,Lee2003,Lan2005,Kaplan2002,Kohler2011} 
and references therein). We have shown, however, that quantum effects prohibit in general the generation of such extremely short pulses 
of radiation and, in some cases, lead to contradictions with the classical expectations \cite{KTK2014a}. The main reason for this is 
that the global phase of the quantum amplitude, $\mathcal{A}_{\mathrm{C},\sigma}(\omega_{\bm{K}})$, nonlinearly depends on the emitted 
photon frequency, $\omega_{\bm{K}}$. This is in contrast to the classical amplitude, $\mathcal{A}_{\mathrm{Th},\sigma}(\omega_{\bm{K}}^{\mathrm{Th}})$, 
the phase of which linearly depends on $\omega^{\mathrm{Th}}_{\bm{K}}$. A nonlinear dependence of the quantum phase leads to strong chirping 
of synthesized pulses of radiation. It appears, however, that the frequency scaling considered in this paper, which is the straightforward 
generalization of the scaling law introduced originally in Refs.~\cite{Heinzl,HSK,Seipt} for laser pulses with slowly changing envelops, 
correctly transforms the Thomson global phase. Namely, after the transformation, it becomes the nonlinear function of $\omega_{\bm{K}}$, 
as it is the case for the Compton global phase \cite{KTK2014b}. One can expect therefore that the frequency scaling law, if applied to 
the classical analysis, can lead to temporal power distributions of emitted radiation comparable to those predicted by the quantum analysis. 
This is the topic of our further discussion presented in the next section.

\section{Temporal power distribution}

The frequency distributions for Compton and Thomson scattering discussed above are not the only ones that can be studied in the context 
of the scaling law. Another aspect of such investigations, in our opinion even more important in light of possible applications, 
is the temporal dependence of power of electromagnetic radiation generated during the interaction of electrons with laser pulses. The aim of this section is 
to show that the meaning of the scaling law can be extended to the time-analysis of generated high-frequency radiation by these two processes.

Analysis of the Li\'enard-Wiechert potentials \cite{Jackson1975,LL2} shows that the Thomson amplitude $\mathcal{A}_{\mathrm{Th},\sigma}(\omega_{\bm{K}})$ 
can be used for the synthesis of the temporal power distribution of generated radiation. Let us take only a part of the frequency spectrum, 
$\omega_{\mathrm{min}}\leqslant\omega_{\bm{K}}\leqslant\omega_{\mathrm{max}}$, by applying for instance a frequency filter, and define the function
\begin{equation}
\tilde{\mathcal{A}}^{(+)}_{\mathrm{Th},\sigma}(\phi_{\mathrm{r}};\omega_{\mathrm{min}},\omega_{\mathrm{max}})
=\int\limits_{\omega_{\mathrm{min}}}^{\omega_{\mathrm{max}}}\mathrm{d}\omega \mathcal{A}_{\mathrm{Th},\sigma}(\omega)
\mathrm{e}^{-\mathrm{i}\omega\phi_{\mathrm{r}}/\omega_{\mathrm{L}}}\, ,
\label{tpd1}
\end{equation}
where the retarded phase $\phi_{\mathrm{r}}$ in the far radiation zone is
\begin{equation}
\phi_{\mathrm{r}}=\omega_{\mathrm{L}}\Bigl(t-\frac{R}{c}\Bigr)\, .
\label{tpd2}
\end{equation}
Here, $R$ is much larger than any displacement of electrons during the interaction with a laser pulse.
The temporal angular power distribution becomes
\begin{align}
\frac{\mathrm{d}^2P_{\mathrm{Th},\sigma}}{\mathrm{d}^2\Omega_{\bm{K}}}\bigl(\phi_{\mathrm{r}};\omega_{\mathrm{min}},& \omega_{\mathrm{max}}\bigr)= 
\frac{e^2}{4\pi^2\varepsilon_0c} \nonumber \\
\times & \bigl[\mathrm{Re} \tilde{\mathcal{A}}^{(+)}_{\mathrm{Th},\sigma}(\phi_{\mathrm{r}};\omega_{\mathrm{min}},\omega_{\mathrm{max}})\bigr]^2 \, ,
\label{tpd3}
\end{align}
where the symbol `$\mathrm{Re}$' means the real part. The corresponding formulas for Compton scattering are the same, except that 
they also depend on the initial and final spin degrees of freedom.

\begin{figure}
\includegraphics[width=7.0cm]{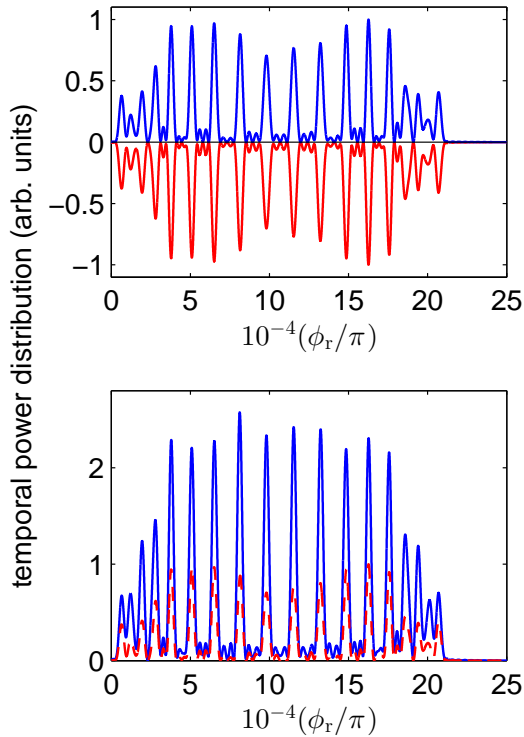}%
\caption{(Color online) Temporal power distribution of electromagnetic radiation generated by Compton and Thomson scattering, 
and synthesize from energy distributions presented in Fig.~\ref{compare2}. In the upper panel, we compare temporal power 
distributions for generated radiation linearly polarized in the scattering plane for the spin-conserved Compton process (blue line) 
and the frequency-scaled Thomson process (red line). Both distributions are normalized to their maximum values. In the lower panel, 
we compare the temporal power distribution for Compton scattering from the upper panel (dashed red line) with the total power 
distribution summed over the final spin and polarization degrees of freedom and averaged over the initial spins (solid blue line). 
We see that spin and polarization effects observed for power distributions are even more pronounced than the ones observed for frequency distributions.
\label{pulse2a}}
\end{figure}

In order to account for the frequency scaling in Thomson scattering, we calculate the complex Thomson amplitude 
$\mathcal{A}_{\mathrm{Th},\sigma}(\omega^{\mathrm{Th}}_{\bm{K}})$, Eqs. \eqref{thom7} and \eqref{thom10}, and scale it such that
\begin{equation}
\mathcal{A}^{\mathrm{scaled}}_{\mathrm{Th},\sigma}(\omega_{\bm{K}})
=\mathcal{A}_{\mathrm{Th},\sigma}\Bigl(\frac{\omega_{\bm{K}}}{1-\omega_{\bm{K}}/\omega_{\mathrm{cut}}}\Bigr)\, .
\label{tpd4}
\end{equation}
This amplitude is then inserted into Eqs. \eqref{tpd1} and \eqref{tpd3} to obtain the frequency-scaled 
temporal power distribution for Thomson scattering.

Fig.~\ref{pulse2a} shows the temporal power distributions synthesized from the frequency distributions presented in Fig.~\ref{compare2}. 
As we see, up to a normalization constant, the frequency-scaled temporal power distribution for Thomson scattering perfectly agrees with the 
corresponding distribution for Compton scattering. Note that, without applying the frequency scaling to the Thomson amplitude, the classical 
electrodynamics predicts generation of much shorter radiation pulses (if for the synthesis such frequencies are used which are 
comparable to the cut-off frequency, $\omega_{\mathrm{cut}}$). The observed agreement proves that not only the square of modulus of frequency-scaled 
Thomson amplitude and Compton amplitude are equal (up to a normalization constant). It proves that dependence of their phases on the frequency 
of emitted radiation are the same up to a constant term (for the Thomson phase, we mean the dependence on the scaled frequency). Again, the results presented 
in Fig.~\ref{pulse2a} show the importance of the spin and polarization degrees of freedom for high-frequency parts of spectra, 
as their contribution to the temporal power distribution can be even more pronounced than for the energy distribution.

In closing, we note that the validity of the frequency scaling law (introduced by Seipt and K\"ampfer~\cite{Seipt} 
for finite but long laser pulses) can be extended not only to arbitrary short laser pulses and to the 'erratic' part of the spectrum, 
but also to the time-domain of quantum and classical theories provided that the electron spin is properly accounted for.
Since the notion of spin is absent in classical theory, one has to realize how to compare both theories in a reliable manner.
Our analysis shows that this is possible only when the Thomson process is compared with the spin-conserved Compton process.
As we also demonstrate, such a comparison makes sense even if the spin-flipping Compton process occurs with a significant probability.

\section{Conclusions}

In this paper, we compared the energy distributions of emitted radiation in nonlinear Compton and Thomson processes by shaped laser pulses.
The presented numerical results were obtained in the framework of quantum and classical electrodynamics, respectively. We observed a typical blue shift of Thomson
spectra with respect to the Compton spectra. However, by employing a respective frequency transformation, we showed that both spectra
start to coincide. Specifically, this concerned the Compton spectra for processes which conserve the electron spin as compared to the Thomson spectra.
Therefore, the importance of spin effects in nonlinear Compton scattering was stressed. In the case when the spin-flipping Compton processes were negligible, 
the frequency transformation was successfully applied to the spin-averaged Compton distributions. One should note, however, that there is a limitation 
on the applicability of the scaling transformation which comes from a sensitivity of classical results to the polarization of emitted radiation. 
This was illustrated when we analyzed angular distributions of emitted radiation.

In closing, we would like to stress that the frequency scaling law introduced in this paper can be successfully applied to Compton and Thomson 
spectra generated by pulses of an arbitrary duration. Moreover, it extends far above a standard validity range of a classical limit [see, Eq. (37)]. 
As was illustrated by numerical examples, our scaling law stays valid even for a high-energy part of the emitted radiation. Finally,
we note that the scaling was previously introduced only in the context of energy distributions of emitted radiation. In this paper, we showed that the frequency transformation
can be also successfully applied to angular distributions and for the time analysis of emitted radiation, which is particularly important in the context of ultra-short pulse generation.

\section*{ACKNOWLEDGMENTS}
This work is supported by the Polish National Science Center 
(NCN) under Grant No.~2012/05/B/ST2/02547.
K.K. gratefully acknowledges the hospitality of the Department of
Physics and Astronomy at the University of Nebraska, Lincoln,
Nebraska, where part of this paper was prepared.
We would also like to thank unknown referees for their suggestions to investigate 
the angular distributions in the context of the frequency scaling law and to study 
the dependence of the total energy of generated radiation on the laser pulse intensity 
for both quantum and classical theories.

\end{document}